\def\extendedversion{}
\documentclass{article}
\usepackage{spconf,amsmath,graphicx,hyperref}

\usepackage{booktabs,array}
\usepackage{multirow}
\usepackage{enumitem}
\usepackage[noend]{algorithmic}
\usepackage{algorithm}
\usepackage{CJKutf8}
\usepackage[nolspread,norspread,nolshrink,norshrink]{cuted}

\newcommand{\tok}[1]{{\footnotesize\texttt{<\textbar#1\textbar>}}}
\newcommand{\onlineappendixurl}{\url{https://adaptduplex.github.io/appendix}}
\ifdefined\extendedversion
  \newcommand{\appendixmention}[1]{Appendix~\ref{#1}}
  \newcommand{\firstappendixmention}[1]{Appendix~\ref{#1}}
\else
  \newcommand{\appendixmention}[1]{Appendix}
  \newcommand{\firstappendixmention}[1]{Appendix\footnote{Appendix: \onlineappendixurl}}
\fi

\title{ADAPTDUPLEX: FROM STATIC TO ADAPTIVE FULL-DUPLEX SPOKEN DIALOGUE}
\name{\shortstack[c]{Zhiyang Zhou, Yingxin Shang, Zhou Wang, Hongwei Cai, Weixu Wang, Shuran Zhou\sthanks{Corresponding author.},\\
Shuofeng Zhao, Wenke Fan, Qingxiang Guo, Dawei Yang, Lin Yang, Yang Song}}
\address{Zuoyebang Education Technology (Beijing) Co., Ltd.}
\ifdefined\zenodoversion
\makeatletter
\def\ps@zenodo{%
  \let\@mkboth\@gobbletwo
  \def\@oddhead{}%
  \def\@evenhead{}%
  \def\@oddfoot{\parbox[t]{\textwidth}{\centering\scriptsize
    This work has been submitted to the IEEE for possible publication.
    Copyright may be transferred without notice, after which this version may no longer be accessible.}}%
  \let\@evenfoot\@oddfoot
}
\makeatother
\fi
\makeatletter
\long\def\@makecaption#1#2{%
  \vskip 4pt
  \setbox\@tempboxa\hbox{#1. #2}%
  \ifdim \wd\@tempboxa >\hsize #1. #2\par \else \hbox to\hsize{\hfil\box\@tempboxa\hfil}\fi}
\makeatother

\begin{document}
\ninept
\ifdefined\appendixonly
  \twocolumn[%
\vspace*{.15em}
{\centering\large\bfseries AdaptDuplex: Online Appendix\par}
\vspace{.8em}
]
\thispagestyle{empty}
\noindent This document reproduces the online appendix of the arXiv extended paper. The appendix begins on the next page and keeps its original section and table numbering.
\vspace{.45em}

\noindent\textbf{Project page.} \url{https://adaptduplex.github.io/}

\vspace{.35em}
\noindent\textbf{Extended paper.} \url{https://arxiv.org/abs/2609.29217}

\vspace{.7em}
\noindent\textbf{Errata and re-evaluation}
\vspace{.35em}

\noindent\textbf{HumDial-FDBench scoring correction}
\vspace{.25em}

\noindent The initial release substituted the wrong reference parameters into the official D-Sco.\ formula. D-Sco.\ is the logarithmic normalization of the aggregated ALL Delay, with $L_{\min}{=}0.624$\,s and $L_{\mathrm{base}}{=}2.531$\,s.

\noindent The corrected values below fix the HumDial scoring parameters used in the initial release. Reported Moshi and Freeze-Omni scores are unchanged published values. Qwen3-Omni+VAD has now been re-run with streaming enabled on HumDial; its results are shown below and are the values reported for Qwen3-Omni+VAD in the HumDial table.

\vspace{.6em}
{\centering\footnotesize
\setlength{\tabcolsep}{3pt}
\begin{tabular}{@{}lccc@{}}
\toprule
System & Initial & Corrected & Streaming re-run \\
& D-Sco./Final & D-Sco./Final & D-Sco./Final \\
\midrule
Qwen3-Omni+VAD & 79.5/66.5 & 23.7/55.3 & 41.9/57.1 \\
MiniCPM-o 4.5 & 94.1/66.7 & 64.0/60.6 & -- \\
DuplexOmni & 86.8/57.3 & 43.8/48.7 & -- \\
AdaptDuplex & 101.8/72.9 & 85.4/69.6 & -- \\
\bottomrule
\end{tabular}\par}

\vspace{3.25ex plus 1ex minus .2ex}
\noindent\textbf{Qwen3-Omni+VAD re-evaluation on FDB}
\vspace{.25em}

\noindent Streaming output was disabled in the initial run. Re-running with streaming enabled can change behavior as well as latency, so both are reported below. These are the values reported for Qwen3-Omni+VAD on FDB-v1 and FDB-v1.5. Behavior follows Table~3.

\vspace{.35em}
{\centering\footnotesize
\setlength{\tabcolsep}{2.5pt}
\begin{tabular}{@{}>{\raggedright\arraybackslash}p{48pt}lcc@{}}
\toprule
\multicolumn{4}{@{}l}{\textit{FDB-v1}} \\
\midrule
Task & Metric & Initial & Streaming re-run \\
\midrule
\multirow{2}{=}{Pause} & Syn.\ TOR$\downarrow$ & 0.891 & 0.993 \\
& Can.\ TOR$\downarrow$ & 0.935 & 0.949 \\
\multirow{3}{=}{Backchannel} & TOR$\downarrow$ & 0.964 & 0.344 \\
& Freq.$\uparrow$ & 0.026 & 0.018 \\
& JSD$\downarrow$ & 0.871 & 0.879 \\
\multirow{2}{=}{Smooth turn} & TOR$\uparrow$ & 0.992 & 0.992 \\
& Lat.$\downarrow$ & 8.454 & 3.358 \\
\multirow{3}{=}{Interruption} & TOR$\uparrow$ & 0.995 & 0.995 \\
& GPT-4o$\uparrow$ & 4.899 & 4.824 \\
& Lat.$\downarrow$ & 9.582 & 3.380 \\
\midrule
\multicolumn{4}{@{}l}{\textit{FDB-v1.5}} \\
\midrule
\multirow{3}{=}{Interruption} & Behavior$\uparrow$ & 0.97 & 0.97 \\
& Stop$\downarrow$ & 0.49 & 0.09 \\
& Resp.$\downarrow$ & 10.28 & 3.70 \\
\multirow{3}{=}{Backchannel} & Behavior$\uparrow$ & 0.00 & 0.00 \\
& Stop$\uparrow$ & -- & -- \\
& Resp.$\downarrow$ & 4.37 & 2.73 \\
\multirow{3}{=}{Talking to others} & Behavior$\uparrow$ & 0.00 & 0.01 \\
& Stop$\uparrow$ & 0.14 & 0.10 \\
& Resp.$\downarrow$ & 6.16 & 3.16 \\
\multirow{3}{=}{Background speech} & Behavior$\uparrow$ & 0.00 & 0.00 \\
& Stop$\uparrow$ & 0.20 & 0.09 \\
& Resp.$\downarrow$ & 6.56 & 3.19 \\
\bottomrule
\end{tabular}\par}
\vspace{.25em}
{\raggedright\footnotesize
FDB-v1.5: Behavior is 1 for Respond/Resume, 0.5 for Uncertain, and 0 otherwise; latencies are in seconds.\par}

\vspace{.8em}
\noindent\textbf{Qwen3-Omni+VAD re-evaluation on HumDial-FDBench}
\vspace{.25em}

{\centering\footnotesize
\setlength{\tabcolsep}{3pt}
\begin{tabular}{@{}llcc@{}}
\toprule
\multicolumn{4}{@{}l}{\textit{HumDial-FDBench}} \\
\midrule
Metric & Split & Initial & Streaming re-run \\
\midrule
\multirow{3}{*}{Int.$\uparrow$ (\%)} & ZH & 93.2 & 91.3 \\
& EN & 84.6 & 80.9 \\
& ALL & 88.8 & 86.1 \\
\multirow{3}{*}{Rej.$\uparrow$ (\%)} & ZH & 36.5 & 36.8 \\
& EN & 38.5 & 34.2 \\
& ALL & 37.6 & 35.6 \\
\multirow{3}{*}{Delay$\downarrow$ (s)} & ZH & 8.514 & 4.670 \\
& EN & 9.536 & 4.924 \\
& ALL & 9.032 & 4.771 \\
D-Sco.$\uparrow$ & ALL & 23.7 & 41.9 \\
Final$\uparrow$ & ALL & 55.3 & 57.1 \\
\bottomrule
\end{tabular}\par}
\vspace{.25em}
{\raggedright\footnotesize
HumDial metrics follow the official protocol. D-Sco.\ is computed from aggregated ALL Delay, and Final $=0.4$ Int. $+0.4$ Rej. $+0.2$ D-Sco.\par}
\clearpage

  \setcounter{table}{4}
  \appendix
  \makeatletter
  \DeclareRobustCommand{\cite}[2][]{#2}
  \makeatother


\section{Protocol Specification}
\label{app:protocol}

\subsection{Sequence grammar}
\label{app:grammar}

\begin{flushleft}
\small\ttfamily\setlength{\parindent}{0pt}%
\setlength{\parskip}{0pt}\raggedright
Episode ::= Prefix (ToolBlock* Unit)+\\
Unit ::= UNIT\_START Environment ASRView Decision UNIT\_END\\
ASRView ::= ASR ASRPayload ASR\_EOS $\mid$ $\epsilon$\\
Decision ::= Listen $\mid$ ContinueSpeak $\mid$ FinalSpeak\\
Listen ::= LISTEN NextWindow CHUNK\_EOS\\
ContinueSpeak ::= SPEAK Action Text NextWindow CHUNK\_EOS\\
FinalSpeak ::= SPEAK Action Text NextWindow TURN\_EOS CHUNK\_EOS\\
Action ::= MEMORY $\mid$ LAUNCH:$N$ $\mid$ CANCEL:$N$\\
\mbox{\phantom{Action ::= }}$\mid$ CANCEL:$N$ LAUNCH:$N$ $\mid$ $\epsilon$,\quad $N\in\{0,1,2,3\}$\\
NextWindow ::= WINDOW\_SHORT $\mid$ WINDOW\_MID $\mid$ WINDOW\_LONG\\
Text ::= TextToken*\\
ToolBlock ::= IM\_START tool ToolPayload IM\_END
\end{flushleft}

\texttt{NextWindow} schedules the following Unit rather than changing the
current one, so both LISTEN and SPEAK end with a window token and
\tok{chunk\_eos}. A speaking
turn is a maximal sequence of SPEAK Units terminated by \texttt{FinalSpeak};
adjacent speaking turns need not be separated by LISTEN. SPEAK and text
presence are orthogonal, so \texttt{Text} may be empty. Tool Blocks occur only
at Unit boundaries, and trailing Tool Blocks without a following Unit are
invalid.

Because every behavioral decision is an explicit token, additive logits bias
changes runtime behavior without modifying weights.
Table~\ref{tab:appendix_logits_bias} lists the three uses.

\begin{table}[!ht]
\caption{Logits-bias controls. Each adds a constant to the named token's logit
at inference.}
\label{tab:appendix_logits_bias}
\centering
\small
\setlength{\tabcolsep}{3pt}
\begin{tabular}{@{}p{0.46\columnwidth}p{0.50\columnwidth}@{}}
\toprule
Bias target & Effect \\
\midrule
\tok{listen} vs.\ \tok{speak} & Shift toward listening or speaking \\
An action token, set to $-\infty$ & Disable that action \\
\tok{window:short}, \tok{window:mid}, \tok{window:long} & Select the next-window duration \\
\bottomrule
\end{tabular}
\end{table}

\subsection{Token roles and supervision}

\begin{table}[!ht]
\caption{Canonical Thinker ranges and training roles. ``Masked'' means that the
range remains in causal context but has no Thinker target.}
\label{tab:appendix_token_roles}
\centering
\small
\setlength{\tabcolsep}{4pt}
\renewcommand{\arraystretch}{1.12}
\begin{tabular}{@{}p{0.30\columnwidth}p{0.46\columnwidth}p{0.16\columnwidth}@{}}
\toprule
Range & Role & Target \\
\midrule
Prefix, Tool Block & Episode/external context & Masked \\
Environment, Unit delimiters & Causal input/boundary & Masked \\
ASR Block & Optional ASR training view & If enabled \\
LISTEN/SPEAK & Interaction control & Valid \\
Action tokens & MEMORY and indexed L2 actions & Valid \\
Assistant text & Reply content & Valid \\
Window token & Schedule following Unit & Valid \\
TURN/CHUNK EOS & Turn/Unit speech boundary & Valid \\
\bottomrule
\end{tabular}
\end{table}

\subsection{State lifetimes}

\begin{table}[!ht]
\caption{State ownership across the three protocol lifetimes.}
\label{tab:appendix_lifetimes}
\centering
\small
\setlength{\tabcolsep}{3.6pt}
\begin{tabular}{@{}p{0.22\columnwidth}p{0.48\columnwidth}p{0.24\columnwidth}@{}}
\toprule
Lifetime & Retained state & Reset boundary \\
\midrule
Episode & Thinker prefix, causal history, committed Units & Episode close \\
Unit & Current environment and canonical decision & Commit to history \\
Speaking turn & Talker KV, text FIFO, codec/PCM rolling state & Final SPEAK \\
\bottomrule
\end{tabular}
\end{table}

Thinker history persists across LISTEN/SPEAK transitions and speaking turns.
Talker state persists across Units only within one speaking turn and is reset
at its terminal SPEAK; a turn reset never clears Thinker episode history.

\subsection{Turn bootstrap and text FIFO}

Each speaking turn owns one Talker row and one FIFO. Before its first acoustic
position, the Talker inserts the five target-free bootstrap conditions in
Table~\ref{tab:appendix_bootstrap}; continuation Units do not repeat them.

\begin{table}[!ht]
\caption{Talker bootstrap at the start of each speaking turn.}
\label{tab:appendix_bootstrap}
\centering
\small
\setlength{\tabcolsep}{3pt}
\begin{tabular}{cll}
\toprule
Position & Text channel & Codec channel \\
\midrule
1 & \tok{tts\_pad} & codec-nothink \\
2 & \tok{tts\_pad} & codec-think-BOS \\
3 & \tok{tts\_pad} & codec-think-EOS \\
4 & \tok{tts\_pad} & speaker identity \\
5 & \tok{tts\_bos} & codec-pad \\
\bottomrule
\end{tabular}
\end{table}

When a SPEAK Unit arrives, its assistant-text tokens are appended in order. A
final Unit also appends \tok{tts\_eos}. Every attended acoustic position pops
one text condition from the left; an empty queue supplies \tok{tts\_pad}.
Unconsumed text therefore carries across Units within the same speaking turn,
but neither the FIFO nor Talker KV state crosses a turn boundary. ASR, control,
action, window, boundary, and Tool-Block tokens are never enqueued. If the
acoustic slots of a turn cannot consume the remaining payload and the final
\tok{tts\_eos}, the turn is rejected rather than built with truncated text.

After the bootstrap, the first Unit combines the first FIFO condition with codec
BOS and predicts its first real frame. A continuation Unit has no synthetic
restart: its predecessor's last real codec frame stays pending, and after the
new Unit's text is enqueued that frame combines with the next FIFO condition to
predict the new Unit's first frame. The pending frame remains a supervised real
frame. No codec BOS, codec PAD, or synthetic boundary is inserted between
ordinary Units.

Ending the turn takes two further steps, and neither produces speech.
Thinker \tok{turn\_eos} only marks the decision; it enqueues \tok{tts\_eos} so
the Talker learns that no text remains. Once the remaining speech has been
consumed, the last real codec frame predicts \tok{codec\_eos}; this is the
same stopping condition in training and runtime, with no separate
\tok{codec\_pad}-conditioned EOS position. The EOS prediction occupies no
acoustic frame and carries no residual-code target, so it supervises where
speech stops rather than what the next frame sounds like. At runtime
\tok{codec\_eos} is masked in every non-final Unit, so speech cannot stop
mid-turn. A final Unit usually holds fewer real frames than its window budget,
since a speaking turn rarely ends on a window boundary, so it may predict
\tok{codec\_eos} before that budget runs out. Only if the budget runs out
first does one extra non-acoustic step check the EOS prediction under the same
last-real-frame condition; anything other than \tok{codec\_eos} is logged and
dropped. The chain is therefore
\tok{turn\_eos}$\rightarrow$\tok{tts\_eos}$\rightarrow$\tok{codec\_eos}.

\subsection{Text-ahead transformation}

For each speaking turn, let $T_k$ be the ordered assistant-text tokens assigned
to its $k$-th Unit. At an internal boundary $k$, Text-ahead samples a lead
$d_k\in[0,d_{\max}]$, with $d_{\max}=2$ in the reported experiments, and moves the first $\min(d_k,|T_{k+1}|)$ tokens of
$T_{k+1}$ to the tail of $T_k$. Transfers use the original per-Unit lists and
occur only between adjacent Units, so applying all boundaries preserves the
flattened turn sequence exactly:
\begin{equation}
  T_1\Vert\cdots\Vert T_K
  = T'_1\Vert\cdots\Vert T'_K.
\end{equation}
The transformation changes only when a token enters the Talker FIFO. It does
not change token order or count, LISTEN/SPEAK and action decisions, codec
targets, or the causal boundary of future environments. Tokens never cross a
speaking-turn boundary. A turn with inconsistent Unit-level supervision masks
is left unchanged, because moving a token would change its training ownership.

The random-ahead schedule draws $(d_k)$ from a bounded random walk whose
transitions increase, hold, decrease, or reset the current lead, with weights
$(3,2,2,1)$; the fixed-ahead schedule sets all internal leads to $d_{\max}$. The
transformation is applied to an eligible turn with probability $p_{\rm TA}=0.8$
and is disabled when $d_{\max}=0$.

\section{Adaptive Window Policy}
\label{app:window}

\subsection{Prediction target and configurations}

At the end of window $k$, the Thinker predicts a three-way token $z_k$ that
schedules window $k{+}1$, using only multimodal context causally available
through the boundary of window $k$. SHORT, MID, and LONG are an ordered
temporal policy, not durations intrinsic to the method. The experimental
configuration instantiates them as in Table~\ref{tab:appendix_window_config};
every duration holds an integer number of 12.5-Hz codec frames. MID initializes
the first window and is the fallback when no earlier rule applies.

\begin{table}[!ht]
\caption{Window configurations used in the experiments.}
\label{tab:appendix_window_config}
\centering
\small
\begin{tabular}{lcc}
\toprule
Token & Duration (s) & Codec frames \\
\midrule
SHORT & 0.48 & 6 \\
MID   & 0.64 & 8 \\
LONG  & 0.96 & 12 \\
\bottomrule
\end{tabular}
\end{table}

\subsection{Weak-label priority rules}

Weak targets are assigned by a deterministic script over dataset metadata:
timestamped acoustic voice-activity boundaries, a complete/incomplete semantic
label on each speaking-to-silence boundary, an interruption/backchannel label
on each overlap, and the ASR transcript with word timestamps. These annotations
are used only to build the target. Training exposes the resulting window token,
and inference predicts it from the Thinker state without the annotations.

The script returns at the first matching rule, so priority is the rule order
and no separate tie-breaking step is needed. The speech-rate test compares the
transcript-derived rate $r$ against a language-specific threshold:
$\tau_{\rm zh}{=}3.5$ Chinese characters/s and $\tau_{\rm en}{=}2.0$ English
words/s, each about $0.7\times$ its conversational rate, so ordinary speech
selects MID and only distinctly slow speech selects LONG.

\begin{algorithm}[!ht]
\caption{Next-window weak target at a window boundary.}
\label{alg:window_target}
\begin{algorithmic}[1]
\renewcommand{\algorithmicif}{\textsc{if}}
\renewcommand{\algorithmicthen}{\textsc{then}}
\renewcommand{\algorithmicelse}{\textsc{else}}
\renewcommand{\algorithmicreturn}{\textsc{return}}
\IF{the window carries an interruption annotation \OR\ the transcript contains a correction marker (\begin{CJK}{UTF8}{gbsn}不对\end{CJK}, \begin{CJK}{UTF8}{gbsn}等等\end{CJK}, \emph{wait}, \emph{actually})}
  \RETURN SHORT
\ENDIF
\IF{a speaking-to-silence boundary falls in the window}
  \RETURN SHORT \textsc{if} its semantic label is complete, \textsc{else} MID
\ENDIF
\IF{the transcript contains a hesitation marker (\begin{CJK}{UTF8}{gbsn}嗯\end{CJK}, \begin{CJK}{UTF8}{gbsn}那个\end{CJK}, \emph{um})}
  \RETURN SHORT
\ENDIF
\IF{silence is sustained}
  \RETURN MID \textsc{if} accumulated silence $<1.0$~s, \textsc{else} LONG
\ENDIF
\IF{a silence-to-speaking transition falls in the window}
  \RETURN MID
\ENDIF
\IF{speech is sustained}
  \RETURN MID \textsc{if} $r>\tau_{\rm lang}$, \textsc{else} LONG
\ENDIF
\IF{the window carries a backchannel annotation}
  \RETURN the preceding window's target
\ENDIF
\RETURN MID
\end{algorithmic}
\end{algorithm}

\section{Tiered Actions and Asynchronous Requests}
\label{app:actions}

\subsection{Action levels}

L0, L1, and L2 are action levels, not token names. L0 is the direct response and
emits no action token. L1 emits \tok{memory}. L2 uses indexed \tok{launch:$N$} and
\tok{cancel:$N$}, $N\in\{0,1,2,3\}$, to control up to four external requests.
Action selection is independent of the LISTEN/SPEAK control: action tokens, when
present, follow \tok{speak} and precede assistant text, as in the grammar of
Appendix~\ref{app:grammar}. Table~\ref{tab:appendix_action_levels} summarizes the
three paths. None of them blocks the current response.

\begin{table}[!ht]
\caption{Action levels and their execution paths.}
\label{tab:appendix_action_levels}
\centering
\footnotesize
\setlength{\tabcolsep}{4pt}
\begin{tabular}{@{}llll@{}}
\toprule
Level & Action token & Execution & Blocking \\
\midrule
L0 & None & Thinker response & No \\
L1 & \tok{memory} & Local backend & No \\
L2 & \tok{launch:$N$}, \tok{cancel:$N$} & External backend & No \\
\bottomrule
\end{tabular}
\end{table}

\subsection{L1 local state}

The minimal L1 backend keeps stable user attributes and preferences in PROFILE
and episodic conversational facts in MEMORY. The parameter-free \tok{memory}
trigger asks the backend to update that state from recent dialogue; it does not
carry field values inside the canonical sequence. The current reply continues
without waiting. A successful update is visible to later Units, and a failed
update neither blocks the current response nor changes the previously committed
state. A richer backend may replace this one provided it keeps the same trigger
and timing.

\subsection{L2 slot lifecycle}

Each of the four slots is independently idle or bound to one in-flight request.
Table~\ref{tab:appendix_l2_transitions} gives the legal per-slot transitions.
NONE is the absence of an L2 action and leaves every slot unchanged. An invalid
action is rejected rather than reinterpreted: launch requires an idle slot and
cancel requires an occupied one.

\begin{table}[!ht]
\caption{Per-slot L2 transitions at a Unit decision.}
\label{tab:appendix_l2_transitions}
\centering
\scriptsize
\setlength{\tabcolsep}{2pt}
\begin{tabular}{@{}ll>{\raggedright\arraybackslash}p{0.30\columnwidth}l@{}}
\toprule
Slot state & Action & Effect & Next state \\
\midrule
Idle & NONE & No external action & Idle \\
Idle & LAUNCH:$N$ & Launch new $g$ & Active($g$) \\
Active($g$) & NONE & Keep request & Active($g$) \\
Active($g$) & CANCEL:$N$ & Cancel $g$ & Idle \\
Active($g$) & CANCEL:$N$; LAUNCH:$N$ & Replace $g$ with $g'$ & Active($g'$) \\
\bottomrule
\end{tabular}
\end{table}

Requests run while full-duplex interaction continues. A cancel invalidates the
slot's active identity before cancellation finishes, so a result that races the
cancellation cannot re-enter as current. Same-slot replacement cancels the old
identity before launching the new one. A result is accepted only when its slot,
request identity, and stage index are still current; duplicates and out-of-order
results are discarded.

\subsection{Return path}

A current request may return one result or an ordered sequence of stages. Each
stage is a complete semantic unit; partial token fragments are never shown to the
Thinker. An accepted stage, whether a valid result or an explicit backend
failure, is serialized as a \texttt{tool}-role Tool Block, given the next stage
index, and inserted at a later Unit boundary. As defined in
Appendix~\ref{app:protocol}, the block is loss-masked but stays in the causal
context, which is how the Thinker distinguishes a reported failure from silence.
Cancelled, stale, and unreturned requests insert nothing. Tool Blocks enter the
Thinker context only and are not queued in the Talker FIFO. No return path
blocks the current response. The protocol specifies no production timeout or
retry; an expired or failed backend is reported, when it is reported, through
the failure block above.

\section{Progressive Training Details}
\label{app:training}

The data-difficulty axis (C1/C2/C3) is orthogonal to the parameter axis
(Thinker-only, Talker-only, joint): only Thinker SFT traverses the curriculum,
while the Talker-only and joint stages each use one fixed data mix.

\subsection{Data generation pipeline}

Training episodes come from an automated multi-agent pipeline. Text-only
dialogue outlines are generated by orchestrator and participant LLM agents,
then rendered into synthetic speech with timestamp alignment. Quality control
combines ASR round-trip verification, overlap sanity checks, and human spot
audits at each curriculum stage. The pipeline is the subject of a companion
study; this section states only what the training setup consumes.

\subsection{Thinker curriculum SFT}
\label{app:thinker_curriculum}

Table~\ref{tab:appendix_curriculum} lists the three stages. Duplex coverage
increases from C1 to C3, and the stage episode sets are disjoint, so a later
stage does not reuse an earlier stage's episodes and no mixing weight is
involved. Each stage runs one epoch at a global batch of 256; a trailing partial
batch counts as one update.

\begin{table}[!ht]
\caption{Thinker curriculum stages.}
\label{tab:appendix_curriculum}
\centering
\setlength{\tabcolsep}{6pt}
\scriptsize
\begin{tabular}{llll}
\toprule
& \multicolumn{1}{c}{C1} & \multicolumn{1}{c}{C2} & \multicolumn{1}{c}{C3} \\
\midrule
Scope & Turn-based & Basic duplex & Full duplex \\
Window sup. & None (fixed MID) & Dynamic & Dynamic \\
Actions & None & None & L1 and L2 \\
Duplex events & None & BC, pause, simple int. & All events \\
Episodes & 63{,}659 & 190{,}976 & 509{,}269 \\
Updates & 249 & 746 & 1{,}990 \\
\bottomrule
\end{tabular}
\end{table}

\subsection{Talker training objective}
\label{app:talker_sft}

Talker-only SFT freezes the Thinker and trains the Talker together with its
residual CodePredictor. Teacher forcing feeds the ground-truth codec sequence,
so each position conditions on the preceding real frames. The main head
predicts group~0 of the current frame; the CodePredictor predicts residual
groups 1--15, giving fifteen residual targets per frame. As specified in
Appendix~\ref{app:protocol}, the final position of a speaking turn predicts
\tok{codec\_eos} from the last real frame and carries no residual target. No
curriculum is used: codec generation conditioned on text and Thinker context is
the same task at every difficulty.

\subsection{Loss weighting and joint SFT}
\label{app:joint_sft}

A per-token-class weight scales the loss numerator; the denominator counts
supervised tokens with no weighting. Both the weighted sum and the count are
reduced across data-parallel workers before the division. Tool Block tokens
are always masked, since they enter from an external source and serve only as
context. Table~\ref{tab:appendix_token_weights} gives the weights.

\begin{table}[!ht]
\caption{Per-token-class loss weights.}
\label{tab:appendix_token_weights}
\centering
\setlength{\tabcolsep}{6pt}
\scriptsize
\begin{tabular}{lcc}
\toprule
Token class & Thinker-only & Joint \\
\midrule
Control (\tok{listen}/\tok{speak}) & 2.0 & 2.0 \\
Window token & 1.5 & 1.5 \\
Action token & 1.5 & 1.5 \\
Assistant text & 1.0 & 1.0 \\
Boundary token & 4.0 & 4.0 \\
ASR view (when enabled) & 1.0 & 1.0 \\
Tool Block & masked & masked \\
\bottomrule
\end{tabular}
\par\vspace{3pt}
\raggedright\scriptsize
The weight scales the loss numerator. The denominator counts supervised tokens without weighting.
\par
\end{table}

Let $S$ be a weighted sum of per-token losses and $N$ the matching unweighted
count. Joint SFT continues on a smaller high-quality set with every component
trainable:
\begin{equation}
\begin{split}
    \mathcal{L}_{\rm thinker} &= S_{\rm thinker}/N_{\rm thinker},\\
    \mathcal{L}_{\rm talker} &= S_{\rm main}/N_{\rm main}
      + 0.3\,S_{\rm res}/N_{\rm res},\\
    \mathcal{L}_{\rm joint} &= \mathcal{L}_{\rm talker}
      + 3\,\mathcal{L}_{\rm thinker}.
\end{split}
\end{equation}

\subsection{Optimization}
\label{app:optimizer}

Table~\ref{tab:appendix_optimizer} collects the optimizer setup. The three
SFT stages share it; GRPO does not inherit it.

\begin{table}[!ht]
\caption{Optimizer setup for the SFT stages and GRPO.}
\label{tab:appendix_optimizer}
\centering
\setlength{\tabcolsep}{6pt}
\scriptsize
\begin{tabular}{lcc}
\toprule
& SFT & GRPO \\
\midrule
Optimizer & \begin{tabular}[c]{@{}c@{}}Adam\\($\beta_1{=}0.9$, $\beta_2{=}0.999$)\end{tabular} & \begin{tabular}[c]{@{}c@{}}Adam\\($\beta_1{=}0.9$, $\beta_2{=}0.95$)\end{tabular} \\
Peak learning rate & $1{\times}10^{-4}$ & $1{\times}10^{-5}$ \\
Global batch size & 256 & 8 \\
Precision & bfloat16 & bfloat16 \\
Hardware & 8$\times$ NVIDIA H100 & 8$\times$ NVIDIA H100 \\
LR schedule & cosine & cosine \\
Warmup ratio & 0.05 & 0 \\
Max sequence length & 32{,}768 & 32{,}768 \\
Gradient clipping & 1.0 & 1.0 \\
Weight decay & 0.01 & 0.1 \\
\bottomrule
\end{tabular}
\end{table}

\subsection{Policy optimization (GRPO)}
\label{app:grpo}

GRPO continues from the joint-SFT checkpoint and updates the interaction policy
without changing the architecture. For each episode, $G$ rollouts are sampled
and the advantage of each is its reward relative to the others in the group, so
no value function is trained.

The reward has three separately evaluated dimensions.
\begin{itemize}[nosep,leftmargin=*]
\item \textbf{Conversational rhythm.} Penalties for late responses, premature
      interruptions, and missed backchannels, computed from the streaming
      timestamps with no human annotation.
\item \textbf{Response quality.} Speech-stream naturalness from a UTMOSv2
      five-fold ensemble, plus text-stream relevance from a reference-free LLM
      judge.
\item \textbf{L2 action appropriateness.} An LLM judge scores whether each
      \tok{launch:$N$} and \tok{cancel:$N$} is justified by the preceding
      context. L0 and L1 are the implicit baselines, so unnecessary and missed
      triggers are both penalized.
\end{itemize}
The composite reward is $\alpha_1 r_{\rm rhythm}+\alpha_2 r_{\rm quality}
+\alpha_3 r_{\rm L2}$. Table~\ref{tab:appendix_grpo} lists the GRPO-specific
quantities.

\begin{table}[!ht]
\caption{GRPO configuration.}
\label{tab:appendix_grpo}
\centering
\small
\begin{tabular}{lc}
\toprule
Quantity & Value \\
\midrule
Rollouts per episode $G$ & 8 \\
Reward weights $\alpha_1,\alpha_2,\alpha_3$ & equal \\
KL coefficient & 0 \\
\bottomrule
\end{tabular}
\end{table}

\section{Experimental Details and Additional Results}
\label{app:experiments}

\subsection{Complete full-duplex results}

Table~\ref{tab:appendix_fdb_full} gives the per-scenario results behind the
aggregated main-text numbers, for every reported training stage and window
policy. Panel (a) is FDB-v1 and panel (b) is FDB-v1.5.

\subsection{Serialization and window-policy execution}

Table~\ref{tab:appendix_execution} reports the token and decoding costs of the
serialization formats and the per-window costs of the window policies,
including the SPEAK/LISTEN decomposition.

\subsection{Text-ahead EOS distributions}

Table~\ref{tab:appendix_text_ahead} breaks the EOS-offset distribution down by
language for the three Text-ahead schedules. The offset bins are $[0,1)$ for
ValidEnd, $[1,3)$ and $[3,5)$ for the intermediate late detail, $[5,16)$ for
Late5+, and $[16,+\infty)$ for NoEOS.

\begin{table}[!t]
\caption{Text-ahead speech quality and EOS-offset distributions.}
\label{tab:appendix_text_ahead}
\begingroup
\ifdefined\tablewide\def\tablewidth{\textwidth}\else\def\tablewidth{\linewidth}\fi
\centering
\scriptsize
\setlength{\tabcolsep}{0pt}
\setlength{\aboverulesep}{0.25ex}
\setlength{\belowrulesep}{0.25ex}
\newcommand{\headbox}[1]{\begin{tabular}[c]{@{}c@{}}#1\end{tabular}}
\makebox[\linewidth][l]{\textbf{(a) Language-specific speech quality}}\par\vspace{2pt}
\begin{tabular}{@{}>{\raggedright\arraybackslash}m{0.230\linewidth}>{\centering\arraybackslash}m{0.090\linewidth}>{\centering\arraybackslash}m{0.090\linewidth}>{\centering\arraybackslash}m{0.205\linewidth}>{\centering\arraybackslash}m{0.205\linewidth}>{\centering\arraybackslash}m{0.090\linewidth}>{\centering\arraybackslash}m{0.090\linewidth}@{}}
\toprule
Group & Lang. & $N$ & UTMOSv2$\uparrow$ & \headbox{Mixed\\UTMOSv2$\uparrow$} & \headbox{CER\\(\%)$\downarrow$} & \headbox{WER\\(\%)$\downarrow$}\\
\midrule
\multirow{2}{*}{Aligned} & ZH & 1,516 & $2.792\!\pm\!.079$ & \multirow{2}{*}{$2.902\!\pm\!.145$} & 10.74 & -- \\
 & EN & 1,514 & $3.012\!\pm\!.107$ &  & -- & 19.10 \\
\midrule
\multirow{2}{*}{Fixed Ahead} & ZH & 1,516 & $3.428\!\pm\!.043$ & \multirow{2}{*}{$3.555\!\pm\!.135$} & 5.38 & -- \\
 & EN & 1,514 & $3.682\!\pm\!.051$ &  & -- & 10.03 \\
\midrule
\multirow{2}{*}{Random Ahead} & ZH & 1,516 & $3.513\!\pm\!.077$ & \multirow{2}{*}{$3.648\!\pm\!.160$} & 4.83 & -- \\
 & EN & 1,514 & $3.783\!\pm\!.092$ &  & -- & 8.93 \\
\bottomrule
\end{tabular}

\vspace{7pt}
\makebox[\linewidth][l]{\textbf{(b) Termination-offset distribution}}\par\vspace{2pt}
\begin{tabular}{@{}>{\raggedright\arraybackslash}m{0.185\linewidth}>{\centering\arraybackslash}m{0.078\linewidth}>{\centering\arraybackslash}m{0.075\linewidth}>{\centering\arraybackslash}m{0.090\linewidth}>{\centering\arraybackslash}m{0.090\linewidth}>{\centering\arraybackslash}m{0.090\linewidth}>{\centering\arraybackslash}m{0.090\linewidth}>{\centering\arraybackslash}m{0.162\linewidth}>{\centering\arraybackslash}m{0.140\linewidth}@{}}
\toprule
Group & Lang. & $N$ & \headbox{$[0,1)$\\(\%)} & \headbox{$[1,3)$\\(\%)} & \headbox{$[3,5)$\\(\%)} & \headbox{$[5,16)$\\(\%)} & \headbox{$[16,+\infty)$\\(\%)} & \headbox{Mean trunc.\\tokens}\\
\midrule
\multirow{2}{*}{Aligned} & ZH & 1,516 & 51.25 & 37.86 & 8.71 & 2.18 & 0.00 & 0.9321\\
 & EN & 1,514 & 48.75 & 22.59 & 14.07 & 13.87 & 0.73 & 1.9478\\
\midrule
\multirow{2}{*}{Fixed Ahead} & ZH & 1,516 & 87.27 & 10.82 & 1.52 & 0.40 & 0.00 & 0.2078\\
 & EN & 1,514 & 82.23 & 9.58 & 4.76 & 3.30 & 0.13 & 0.5436\\
\midrule
\multirow{2}{*}{Random Ahead} & ZH & 1,516 & 90.04 & 8.71 & 1.12 & 0.13 & 0.00 & 0.1563\\
 & EN & 1,514 & 82.69 & 9.78 & 4.62 & 2.84 & 0.07 & 0.4808\\
\bottomrule
\end{tabular}
\par\vspace{3pt}
\raggedright\footnotesize
Panel (a): $N$ counts evaluated utterances. UTMOSv2 is the mean $\pm$ standard deviation; CER applies to ZH and WER to EN, so the other cell is --, and Mixed UTMOSv2 pools both. Panel (b): entries are percentages of the language-specific turns, and mean trunc. tokens uses the same subset.
\par
\endgroup

\end{table}

\subsection{L1 and L2 diagnostics}

Action prediction and information utilization share the 1{,}000 evaluation
episodes but score different views. Action prediction expands the episodes into
156{,}376 jointly labelled L1/L2 decision points and uses gold prefixes.
Information utilization instead supplies gold memory or Tool-Block context, so
it scores the use of information rather than the decision to request it.
Table~\ref{tab:appendix_action_prediction} reports the action decisions and
Table~\ref{tab:appendix_information_utilization} the utilization results.

\begin{table}[!t]
\caption{L1 and L2 action prediction.}
\label{tab:appendix_action_prediction}
\begingroup
\ifdefined\tablewide\def\tablewidth{\textwidth}\else\def\tablewidth{\linewidth}\fi
\newcommand{\shead}[1]{\begin{tabular}[c]{@{}c@{}}#1\end{tabular}}
\centering
\scriptsize
\setlength{\tabcolsep}{0pt}
\setlength{\aboverulesep}{0.25ex}
\setlength{\belowrulesep}{0.25ex}
\begin{tabular}{@{}>{\raggedright\arraybackslash}m{0.20\tablewidth}*{7}{>{\centering\arraybackslash}m{0.1142857\tablewidth}}@{}}
\toprule
\multicolumn{8}{@{}l}{\textbf{(a) L1 action prediction}}\\
\midrule
Scope & \shead{Decision\\points} & TP & FP & FN & Prec.$\uparrow$ & Rec.$\uparrow$ & F1$\uparrow$\\
\midrule
Overall & 156,376 & 1,397 & 329 & 341 & 80.94 & 80.38 & 80.66\\
Multi & 51,636 & 633 & 171 & 197 & 78.73 & 76.27 & 77.48\\
\midrule
\end{tabular}\par\nointerlineskip
\begin{tabular}{@{}>{\raggedright\arraybackslash}m{0.17\tablewidth}>{\raggedright\arraybackslash}m{0.18\tablewidth}>{\centering\arraybackslash}m{0.10\tablewidth}>{\centering\arraybackslash}m{0.11\tablewidth}*{2}{>{\centering\arraybackslash}m{0.18\tablewidth}}>{\centering\arraybackslash}m{0.08\tablewidth}@{}}
\multicolumn{7}{@{}l}{\textbf{(b) L2 action prediction}}\\
\midrule
Occupancy & Gold action & \shead{Decision\\points} & EM$\uparrow$ & \shead{Launch\\P/R/F1$\uparrow$} & \shead{Cancel\\P/R/F1$\uparrow$} & \shead{Invalid\\$\downarrow$}\\
\midrule
Empty & Keep & 34,761 & 92.5 & -- & -- & 1.72\\
Empty & Launch & 7,062 & 78.3 & 82/77/79 & -- & 0.30\\
Single & Keep & 40,238 & 86.7 & -- & -- & 5.15\\
Single & Launch & 8,917 & 70.2 & 78/72/75 & -- & 0.84\\
Single & Cancel & 9,846 & 58.4 & -- & 70/61/65 & 1.02\\
Single & C+L & 3,916 & 49.3 & 74/68/71 & 66/57/61 & 1.65\\
Multi (2--3) & Keep & 24,713 & 78.1 & -- & -- & 8.11\\
Multi (2--3) & Launch & 5,842 & 61.1 & 71/64/67 & -- & 5.56\\
Multi (2--3) & Cancel & 9,276 & 52.8 & -- & 58/46/51 & 6.71\\
Multi (2--3) & C+L & 5,453 & 37.4 & 66/58/62 & 52/40/45 & 7.23\\
Full (4) & Keep & 3,411 & 74.4 & -- & -- & 15.09\\
Full (4) & Cancel & 2,116 & 48.6 & -- & 54/42/47 & 6.35\\
Full (4) & C+L & 825 & 33.9 & 61/53/57 & 48/35/41 & 13.28\\
\midrule
\end{tabular}\par\nointerlineskip
\begin{tabular}{@{}>{\raggedright\arraybackslash}m{0.20\tablewidth}*{5}{>{\centering\arraybackslash}m{0.16\tablewidth}}@{}}
\multicolumn{6}{@{}l}{\textbf{(c) L2 invalid-action breakdown}}\\
\midrule
Scope & \shead{Other\\occupied\\launch} & \shead{Other idle\\cancel} & \shead{Duplicate\\launch} & \shead{Duplicate\\cancel} & Total\\
\midrule
Overall & 1,000 & 570 & 410 & 260 & 2,240\\
Multi & 860 & 460 & 350 & 215 & 1,885\\
Full & 150 & 75 & 65 & 35 & 325\\
\bottomrule
\end{tabular}
\par\vspace{3pt}
\raggedright\footnotesize
The shared evaluation set contains 1,000 episodes and 156,376 decision points. Precision, recall, F1, EM, and Invalid are percentages; decision points, TP, FP, FN, and panel (c) are counts. C+L denotes cancel+launch. P/R/F1 match the action token and slot; EM executes the sequence in order and permits permutation among initially idle slots. Invalid tokens count as false positives and make EM zero. Panel (b) uses Multi (2--3); Multi in panels (a,c) means at least two in-flight requests and includes Full. In panel (c) each invalid token gets one category, with duplicate errors preceding occupied-launch or idle-cancel errors and Full occupying all four slots.
\par
\endgroup

\end{table}

  \begin{table*}[!t]
  \def\tablewide{}%
  \caption{Complete FDB-v1 turn-taking and FDB-v1.5 overlap-handling results.}%
  \label{tab:appendix_fdb_full}%
  \begingroup
\ifdefined\tablewide\def\columnwidth{\textwidth}\else\def\columnwidth{\linewidth}\fi
\newcommand{\headbox}[1]{\begin{tabular}[c]{@{}c@{}}#1\end{tabular}}
\newcommand{\fixedhead}[1]{\makebox[0.055\columnwidth][c]{\headbox{#1}}}
\centering
\footnotesize
\setlength{\tabcolsep}{2pt}
\setlength{\aboverulesep}{0.25ex}
\setlength{\belowrulesep}{0.25ex}
\begin{tabular}{@{}>{\raggedright\arraybackslash}p{0.105\columnwidth}>{\raggedright\arraybackslash}p{0.085\columnwidth}>{\centering\arraybackslash}p{0.055\columnwidth}>{\centering\arraybackslash}p{0.055\columnwidth}>{\centering\arraybackslash}p{0.055\columnwidth}>{\centering\arraybackslash}p{0.060\columnwidth}>{\centering\arraybackslash}p{0.055\columnwidth}>{\centering\arraybackslash}p{0.055\columnwidth}>{\centering\arraybackslash}p{0.055\columnwidth}>{\centering\arraybackslash}p{0.055\columnwidth}>{\centering\arraybackslash}p{0.055\columnwidth}>{\centering\arraybackslash}p{0.055\columnwidth}>{\centering\arraybackslash}p{0.055\columnwidth}>{\centering\arraybackslash}p{0.055\columnwidth}@{}}
\toprule
\multicolumn{1}{@{}l}{\makebox[0.105\columnwidth][l]{Task}} &
\multicolumn{1}{l}{\makebox[0.085\columnwidth][l]{Metric}} &
\multicolumn{12}{@{}c@{}}{\begin{tabular}[c]{*{12}{c}@{}}
\multicolumn{5}{c}{Reference systems} & \multicolumn{7}{c}{AdaptDuplex training} \\
\cmidrule(lr){1-5}\cmidrule(l){6-12}
\multicolumn{1}{c}{\fixedhead{Moshi$^\dagger$}} &
\multicolumn{1}{c}{\fixedhead{Freeze-\\Omni$^\dagger$}} &
\multicolumn{1}{c}{\fixedhead{Qwen3-\\Omni\\+VAD}} &
\multicolumn{1}{c}{\makebox[0.060\columnwidth][c]{\headbox{MiniCPM\\-o 4.5}}} &
\multicolumn{1}{c}{\fixedhead{Duplex\\Omni}} &
\multicolumn{1}{c}{\fixedhead{Mixed\\SFT}} &
\multicolumn{4}{@{}c@{}}{\begin{tabular}[c]{cccc}
\multicolumn{4}{c}{Curriculum} \\
\cmidrule(lr){1-4}
\fixedhead{Fixed\\0.48 s} & \fixedhead{Fixed\\0.64 s} & \fixedhead{Fixed\\0.96 s} & \fixedhead{Dynamic}
\end{tabular}} &
\multicolumn{1}{c}{\fixedhead{+ Joint\\SFT}} &
\multicolumn{1}{c@{}}{\fixedhead{+ RL\\(GRPO)}} \\
\end{tabular}}\\
\midrule
\multicolumn{14}{@{}l}{\textbf{(a) FDB-v1: turn-taking}} \\
\midrule
\multirow{2}{*}{Pause}
& Syn. TOR$\downarrow$ & 0.985 & 0.642 & 0.993 & 0.270 & 0.116 &  & 0.131 & 0.130 & 0.125 & 0.128 & 0.126 & 0.123 \\
& Can. TOR$\downarrow$ & 0.980 & 0.481 & 0.949 & 0.463 & 0.097 &  & 0.431 & 0.406 & 0.378 & 0.299 & 0.291 & 0.284 \\
\midrule
\multirow{3}{*}{\shortstack[l]{Back-\\channel}}
& TOR$\downarrow$ & 1.000 & 0.636 & 0.344 & 0.273 & 0.491 &  & 0.346 & 0.335 & 0.327 & 0.334 & 0.331 & 0.327 \\
& Freq.$\uparrow$ & 0.001 & 0.001 & 0.018 & 0.020 & 0.034 &  & 0.041 & 0.040 & 0.039 & 0.040 & 0.040 & 0.041 \\
& JSD$\downarrow$ & 0.957 & 0.997 & 0.879 & 0.902 & 0.847 &  & 0.824 & 0.823 & 0.818 & 0.821 & 0.818 & 0.815 \\
\midrule
\multirow{2}{*}{\shortstack[l]{Smooth\\turn}}
& TOR$\uparrow$ & 0.941 & 0.336 & 0.992 & 0.882 & 0.193 &  & 0.992 & 0.980 & 0.958 & 0.982 & 0.984 & 0.985 \\
& Resp. lat.$\downarrow$ & 0.265 & 0.953 & 3.358 & 1.222 & 0.000$^\ddagger$ &  & 0.052 & 0.079 & 0.112 & 0.085 & 0.079 & 0.076 \\
\midrule
\multirow{3}{*}{\shortstack[l]{Inter-\\ruption}}
& TOR$\uparrow$ & 1.000 & 0.867 & 0.995 & 0.985 & 0.746 &  & 0.985 & 0.983 & 0.985 & 0.985 & 0.986 & 0.986 \\
& GPT-4o$\uparrow$ & 0.765 & 3.615 & 4.824 & 4.497 & 4.222 &  & 4.168 & 4.174 & 4.216 & 4.512 & 4.543 & 4.570 \\
& Resp. lat.$\downarrow$ & 0.257 & 1.409 & 3.380 & 1.408 & 0.158 &  & 0.104 & 0.163 & 0.234 & 0.131 & 0.119 & 0.113 \\
\midrule
\multicolumn{14}{@{}l}{\textbf{(b) FDB-v1.5: overlap handling}} \\
\midrule
\multirow{6}{*}{Interruption}
& Respond$\uparrow$   & 0.50 & 0.72 & 0.97 & 0.82 & 0.67 & 0.62 & 0.82 & 0.78 & 0.76 & 0.79 & 0.83 & 0.85 \\
& Resume$\downarrow$  & 0.26 & 0.12 & 0.00 & 0.10 & 0.12 & 0.13 & 0.10 & 0.11 & 0.12 & 0.11 & 0.09 & 0.08 \\
& Uncertain$\downarrow$ & 0.00 & 0.03 & 0.03 & 0.01 & 0.00 & 0.01 & 0.02 & 0.00 & 0.00 & 0.01 & 0.00 & 0.00 \\
& Unknown$\downarrow$ & 0.25 & 0.13 & 0.01 & 0.07 & 0.22 & 0.24 & 0.08 & 0.11 & 0.12 & 0.09 & 0.08 & 0.07 \\
& Stop lat.$\downarrow$ & 1.16 & 1.42 & 0.09 & 2.49 & 2.01 & 2.86 & 1.19 & 1.81 & 2.61 & 1.32 & 1.20 & 1.08 \\
& Resp. lat.$\downarrow$ & 1.47 & 1.35 & 3.70 & 1.86 & 0.36 & 1.25 & 0.33 & 0.42 & 0.58 & 0.34 & 0.31 & 0.27 \\
\midrule
\multirow{6}{*}{Backchannel}
& Respond$\downarrow$ & 0.02 & 0.07 & 0.34 & 0.01 & 0.01 & 0.00 & 0.01 & 0.01 & 0.01 & 0.01 & 0.00 & 0.00 \\
& Resume$\uparrow$ & 0.06 & 0.80 & 0.00 & 0.46 & 0.28 & 0.41 & 0.52 & 0.54 & 0.56 & 0.55 & 0.55 & 0.57 \\
& Uncertain$\downarrow$ & 0.00 & 0.02 & 0.27 & 0.00 & 0.02 & 0.01 & 0.01 & 0.00 & 0.00 & 0.00 & 0.00 & 0.01 \\
& Unknown$\downarrow$ & 0.92 & 0.11 & 0.40 & 0.53 & 0.69 & 0.58 & 0.46 & 0.45 & 0.43 & 0.44 & 0.45 & 0.42 \\
& Stop lat.$\uparrow$ & 0.42 & 0.66 & -- & 0.76 & 0.80 & 0.86 & 0.82 & 0.84 & 0.85 & 0.84 & 0.86 & 0.89 \\
& Resp. lat.$\downarrow$ & 3.00 & 2.16 & 2.73 & 2.34 & 0.82 & 1.77 & 0.32 & 0.66 & 1.14 & 0.43 & 0.32 & 0.31 \\
\midrule
\multirow{6}{*}{\shortstack[l]{Talking\\to others}}
& Respond$\downarrow$ & 0.20 & 0.58 & 0.96 & 0.64 & 0.15 & 0.47 & 0.53 & 0.51 & 0.50 & 0.50 & 0.47 & 0.46 \\
& Resume$\uparrow$ & 0.19 & 0.25 & 0.00 & 0.20 & 0.36 & 0.26 & 0.29 & 0.30 & 0.33 & 0.33 & 0.35 & 0.41 \\
& Uncertain$\uparrow$ & 0.02 & 0.00 & 0.01 & 0.02 & 0.01 & 0.03 & 0.04 & 0.05 & 0.01 & 0.03 & 0.04 & 0.05 \\
& Unknown$\downarrow$ & 0.59 & 0.15 & 0.03 & 0.14 & 0.47 & 0.24 & 0.14 & 0.14 & 0.16 & 0.14 & 0.14 & 0.08 \\
& Stop lat.$\uparrow$ & 0.87 & 1.39 & 0.10 & 1.71 & 1.87 & 1.89 & 1.88 & 1.90 & 1.95 & 1.89 & 1.91 & 1.87 \\
& Resp. lat.$\downarrow$ & 2.38 & 1.32 & 3.16 & 1.85 & 1.95 & 1.53 & 0.32 & 0.32 & 0.34 & 0.33 & 0.32 & 0.29 \\
\midrule
\multirow{6}{*}{\shortstack[l]{Background\\speech}}
& Respond$\downarrow$ & 0.21 & 0.63 & 0.98 & 0.43 & 0.05 & 0.47 & 0.50 & 0.44 & 0.37 & 0.42 & 0.40 & 0.36 \\
& Resume$\uparrow$ & 0.07 & 0.25 & 0.00 & 0.36 & 0.07 & 0.23 & 0.33 & 0.42 & 0.46 & 0.42 & 0.45 & 0.42 \\
& Uncertain$\uparrow$ & 0.01 & 0.01 & 0.00 & 0.02 & 0.00 & 0.01 & 0.02 & 0.01 & 0.00 & 0.01 & 0.02 & 0.06 \\
& Unknown$\downarrow$ & 0.71 & 0.11 & 0.02 & 0.19 & 0.89 & 0.29 & 0.15 & 0.13 & 0.17 & 0.15 & 0.13 & 0.16 \\
& Stop lat.$\uparrow$ & 0.54 & 0.98 & 0.09 & 1.46 & 1.50 & 1.46 & 1.44 & 1.47 & 1.59 & 1.51 & 1.56 & 1.60 \\
& Resp. lat.$\downarrow$ & 1.62 & 1.60 & 3.19 & 1.61 & 0.74 & 2.07 & 0.30 & 0.39 & 0.59 & 0.36 & 0.33 & 0.31 \\
\bottomrule
\end{tabular}
\par\vspace{3pt}
\raggedright\footnotesize
(a): Freq. in events/s; GPT-4o scores 0--5; latencies in seconds. GPT-4o score and all latencies conditional on TO=1. Mixed SFT is not evaluated on FDB-v1. $^\ddagger$Negative smooth-turn latencies clipped to zero; excluded from best-value highlighting. $\dagger$From FDB-v1~\cite{lin2025fullduplexbenchbenchmarkevaluatefullduplex} Table III; latencies not hardware-matched.\\
(b): Respond/Resume/Uncertain/Unknown are rates (0--1); Stop/Response are latencies (s). --: no qualifying event. $\dagger$From FDB-v1.5~\cite{lin2026fullduplexbenchv15evaluatingoverlap} Table 2; latencies are not hardware-matched.
\par
\endgroup
  \end{table*}

  \begin{table*}[!t]
  \def\tablewide{}%
  \caption{Execution statistics for serialization formats and window policies.}%
  \label{tab:appendix_execution}%
  \begingroup
\ifdefined\tablewide\def\columnwidth{\textwidth}\else\def\columnwidth{\linewidth}\fi
\newcommand{\headbox}[1]{\begin{tabular}[c]{@{}c@{}}#1\end{tabular}}
\centering
\footnotesize
\setlength{\tabcolsep}{0pt}
\setlength{\aboverulesep}{0.25ex}
\setlength{\belowrulesep}{0.25ex}
\begin{tabular}{@{}>{\raggedright\arraybackslash}m{0.10\columnwidth}>{\raggedright\arraybackslash}m{0.12\columnwidth}>{\raggedright\arraybackslash}m{0.08\columnwidth}>{\centering\arraybackslash}m{0.075\columnwidth}>{\centering\arraybackslash}m{0.085\columnwidth}>{\centering\arraybackslash}m{0.10\columnwidth}>{\centering\arraybackslash}m{0.09\columnwidth}*{5}{>{\centering\arraybackslash}m{0.07\columnwidth}}@{}}
\toprule
Format & Window policy & Group & \multicolumn{9}{@{}c@{}}{%
\begin{tabular}[c]{@{}>{\centering\arraybackslash}m{0.075\columnwidth}>{\centering\arraybackslash}m{0.085\columnwidth}>{\centering\arraybackslash}m{0.10\columnwidth}>{\centering\arraybackslash}m{0.09\columnwidth}*{5}{>{\centering\arraybackslash}m{0.07\columnwidth}}@{}}
\multicolumn{2}{c}{Workload} & \multicolumn{2}{c}{Comp. Cost} & \multicolumn{5}{c}{Exec. Time (ms)}\\
\cmidrule(lr){1-2}\cmidrule(lr){3-4}\cmidrule(l){5-9}
Windows & \headbox{Input\\(min)} & \headbox{GPU time\\(s)} & \headbox{GPU-s/min\\$\downarrow$} & Mean & P50$\downarrow$ & P90$\downarrow$ & P95$\downarrow$ & P99$\downarrow$\\
\end{tabular}}\\
\midrule
\multirow{3}{=}{\centering JSON-style} & \multirow{3}{=}{\centering Fixed 0.48 s} & All & 3,724 & 29.792 & 2,833.331 & 95.104 & 760.8 & 789.4 & 957.6 & 1,018.7 & 1,164.2\\
& & SPEAK & 2,503 & 20.024 & 2,177.410 & 108.740 & 870.0 & 844.8 & 1,041.2 & 1,116.5 & 1,269.3\\
& & LISTEN & 1,221 & 9.768 & 655.921 & 67.150 & 537.2 & 529.6 & 658.4 & 711.8 & 879.4\\
\midrule
\multirow{12}{=}{\centering Canonical} & \multirow{3}{=}{\centering Fixed 0.48 s} & All & 3,724 & 29.792 & 1,439.586 & 48.321 & 386.6 & 418.5 & 491.2 & 565.9 & 672.2\\
& & SPEAK & 2,503 & 20.024 & 1,127.973 & 56.331 & 450.6 & 433.1 & 527.9 & 605.5 & 689.9\\
& & LISTEN & 1,221 & 9.768 & 311.613 & 31.901 & 255.2 & 254.3 & 316.7 & 359.7 & 506.7\\
\cmidrule(l){2-12}
& \multirow{3}{=}{\centering Fixed 0.64 s} & All & 2,790 & 29.760 & 1,223.506 & 41.112 & 438.5 & 480.1 & 581.7 & 638.5 & 741.8\\
& & SPEAK & 1,901 & 20.277 & 985.681 & 48.610 & 518.5 & 500.5 & 614.6 & 678.5 & 768.6\\
& & LISTEN & 889 & 9.483 & 237.825 & 25.080 & 267.5 & 265.5 & 338.9 & 384.0 & 538.3\\
\cmidrule(l){2-12}
& \multirow{3}{=}{\centering Fixed 0.96 s} & All & 1,856 & 29.696 & 993.492 & 33.455 & 535.3 & 595.3 & 714.6 & 773.3 & 868.8\\
& & SPEAK & 1,298 & 20.768 & 827.248 & 39.833 & 637.3 & 619.1 & 745.8 & 802.3 & 901.2\\
& & LISTEN & 558 & 8.928 & 166.244 & 18.621 & 297.9 & 293.7 & 381.1 & 440.8 & 613.1\\
\cmidrule(l){2-12}
& \multirow{3}{=}{\centering Dynamic} & All & 2,430 & 29.712 & 1,304.880 & 43.920 & 465.2 & 492.6 & 603.8 & 665.1 & 781.4\\
& & SPEAK & 1,660 & 20.064 & 1,056.420 & 52.615 & 548.7 & 541.3 & 658.9 & 721.6 & 836.5\\
& & LISTEN & 770 & 9.648 & 248.460 & 25.751 & 285.7 & 286.4 & 354.2 & 399.8 & 566.9\\
\bottomrule
\end{tabular}
\par\vspace{3pt}
\raggedright\footnotesize
Input is the duration attributed to each group. Cost = GPU seconds / input minutes; ALL aggregates SPEAK and LISTEN. Exec. time is distinct from interaction latency.
\par
\endgroup
  \end{table*}

  \begin{table*}[!t]
  \def\tablewide{}%
  \caption{L1 and L2 information utilization.}%
  \label{tab:appendix_information_utilization}%
  \begingroup
\ifdefined\tablewide\def\tablewidth{\textwidth}\else\def\tablewidth{\linewidth}\fi
\newcommand{\shead}[1]{\begin{tabular}[c]{@{}c@{}}#1\end{tabular}}
\centering
\footnotesize
\setlength{\tabcolsep}{0pt}
\setlength{\aboverulesep}{0.25ex}
\setlength{\belowrulesep}{0.25ex}
\begin{tabular}{@{}>{\raggedright\arraybackslash}p{0.18\tablewidth}>{\centering\arraybackslash}p{0.13\tablewidth}>{\centering\arraybackslash}p{0.12\tablewidth}>{\centering\arraybackslash}p{0.13\tablewidth}>{\centering\arraybackslash}p{0.12\tablewidth}>{\centering\arraybackslash}p{0.16\tablewidth}>{\centering\arraybackslash}p{0.16\tablewidth}@{}}
\toprule
\multicolumn{7}{@{}l}{\textbf{(a) L1 memory utilization}}\\
\midrule
Condition & \shead{Fact\\correct/N} & \shead{Fact\\Acc.$\uparrow$} & \shead{Pref.\\correct/N} & \shead{Pref.\\Acc.$\uparrow$} & \shead{FalseUpd.\\errors/N} & \shead{FalseUpd.\\$\downarrow$}\\
\midrule
Rolling only & 41/400 & 10.25 & 67/400 & 16.75 & -- & --\\
Gold P + M & 251/400 & 62.75 & 237/400 & 59.25 & 17/200 & 8.50\\
\midrule
\end{tabular}\par\nointerlineskip
\begin{tabular}{@{}>{\raggedright\arraybackslash}p{0.18\tablewidth}>{\centering\arraybackslash}p{0.10\tablewidth}*{4}{>{\centering\arraybackslash}p{0.18\tablewidth}}@{}}
\multicolumn{6}{@{}l}{\textbf{(b) L2 external-reasoning utilization}}\\
\midrule
Condition & N & Correct & \shead{Behavior\\Acc.$\uparrow$} & \shead{Missing\\target} & \shead{Forbidden\\value}\\
\midrule
Single & 200 & 173 & 86.50 & 27 & 0\\
Bind & 200 & 147 & 73.50 & 31 & 27\\
Compose & 200 & 137 & 68.50 & 63 & 8\\
Fail & 200 & 163 & 81.50 & 21 & 20\\
None & 200 & 161 & 80.50 & -- & 39\\
\midrule
\end{tabular}\par\nointerlineskip
\begin{tabular}{@{}>{\raggedright\arraybackslash}p{0.20\tablewidth}>{\centering\arraybackslash}p{0.16\tablewidth}*{3}{>{\centering\arraybackslash}p{0.213333\tablewidth}}@{}}
\multicolumn{5}{@{}l}{\textbf{(c) Bind counterfactual intervention}}\\
\midrule
Condition & Pairs & \shead{Original\\Acc.$\uparrow$} & \shead{Swapped\\Acc.$\uparrow$} & \shead{Pair\\Acc.$\uparrow$}\\
\midrule
Bind & 200 & 73.50 & 70.50 & 64.00\\
\bottomrule
\end{tabular}
\par\vspace{3pt}
\raggedright\footnotesize
All accuracies and FalseUpd. are percentages; other entries are counts. Panel (a) uses 1,000 instances: 400 fact, 400 preference, and 200 no-update cases; the two context conditions are paired on the fact and preference instances. Panel (b) uses 1,000 scenarios, 200 per condition. Behavior Acc. equals one iff all target canaries and no forbidden canaries appear. Missing-target and forbidden-value errors may co-occur. Panel (c) is a paired intervention on the 200 Bind cases: A/B payloads are exchanged while the dialogue, query, and request identities remain fixed, Pair Acc. requires both the original and swapped members to be correct.
\par
\endgroup
  \end{table*}

\else
\maketitle
\ifdefined\zenodoversion
\thispagestyle{zenodo}
\fi
\begin{abstract}
Full-duplex spoken dialogue requires simultaneous listening and speaking at sub-second latency, under conversational timing and cognitive demands that change moment to moment.
Yet current models mostly impose static operating points, lacking a systematic mechanism for adaptive decisions.
We present AdaptDuplex, which extends Qwen3-Omni with such a mechanism, co-designed across three layers.
A compact token-level protocol represents every window as a canonical sequence, trains dual-stream alignment through a bounded text lead over speech, and exposes every behavioral decision as an explicit token for training-free runtime control via logits bias.
Adaptive mechanisms dynamically predict among discrete window durations and augment direct response as needed with non-blocking cognitive consolidation and multi-flight external reasoning.
A progressive pipeline introduces these behaviors through a three-stage Thinker curriculum, then Talker-only and joint SFT, with GRPO as a preliminary increment.
On Full-Duplex-Bench v1 and v1.5, AdaptDuplex outperforms DuplexOmni on 18 of 21 comparable turn-taking, overlap-behavior, and timing metrics and MiniCPM-o~4.5 on 19 of 22, with gains in both interaction decisions and response timing.
On human-recorded HumDial-FDBench, it attains the top Final score (69.6) of the compared duplex models.

\end{abstract}
\begin{keywords}
full-duplex spoken dialogue, real-time, token-level protocol, adaptive interaction, curriculum learning
\end{keywords}
\section{Introduction}
\label{sec:introduction}
Human conversation is inherently full-duplex: listeners acknowledge, interrupt, and revise while the other party is still speaking.
Replicating this requires a spoken dialogue model that listens continuously while generating, coordinates overlapping speech, and responds within sub-second time scales as conversational timing and cognitive demands shift from moment to moment.
Dedicated duplex models have established the feasibility of such interaction~\cite{defossez2024moshispeechtextfoundationmodel,ma2024languagemodellistenspeaking,zhang2025omniflattenendtoendgptmodel,yu2024salmonnomnicodecfreellmfullduplex,fang2026baylingduplexnativefullduplexspeech}, and strong omni-models provide a capable pretrained starting point for multimodal understanding and generation; yet a full-duplex model that adapts as these demands change remains hard to obtain.

Existing duplex models still impose a static operating point: window duration and available actions are hard-wired at design time, and the protocol is either too heavy to grow with each added decision or too sparse to carry a richer policy.
DuplexOmni~\cite{huang2026duplexomnirealtimelisteningseeing} and MiniCPM-o~4.5~\cite{cui2026minicpmo45realtimefullduplex} both fix the window duration; the former admits only single-flight external reasoning without intermediate cognitive consolidation, and the latter provides only binary interaction control.
Adaptive full-duplex interaction therefore requires a real-time stream of context-dependent decisions: temporal granularity must remain selectable and available actions composable at runtime, rather than hard-wired into the protocol and training setup.
The text--speech relation is likewise fixed: frame-synchronous models such as Moshi~\cite{defossez2024moshispeechtextfoundationmodel} act on every 80-ms frame with a time-aligned text stream, and streaming TTS interleaves text and speech tokens at a fixed ratio~\cite{du2024cosyvoice2scalablestreaming}, so neither lets the text lead over speech vary within a bound.

Introducing these decisions into a pretrained omni-model presents a further challenge.
Prior work has separately explored state tokens, latent cognition, and policy optimization for interaction control~\cite{wang2024freezeomnismartlowlatency,wu2026silentthoughtmodelinginternal,li2026decouplingconversationaldynamicsfullduplex}, but exposing a pretrained model to protocol adherence, interaction control, and speech generation all at once entangles their learning objectives, and staging training by data scale, quality, trainable modules, or half- to full-duplex mode~\cite{zhang2025omniflattenendtoendgptmodel} does not order individual duplex behaviors by difficulty.

We initialize AdaptDuplex from Qwen3-Omni~\cite{xu2025qwen3omnitechnicalreport}, building on its pretrained multimodal capabilities rather than learning duplex competence from scratch, and co-design three layers---a compact canonical sequence protocol, adaptive interaction mechanisms, and a progressive training pipeline---around one principle: every behavioral decision is an explicit token (Fig.~\ref{fig:overview}).

Our contributions are threefold:
\begin{enumerate}[nosep,leftmargin=*]
    \item We formalize a compact canonical sequence protocol for full-duplex interaction. It reduces structural decoding overhead, trains dual-stream alignment through a bounded text lead over speech, and exposes interaction decisions to training-free runtime control through logits bias.
    \item We propose adaptive interaction mechanisms along two axes: dynamically predicting among discrete window durations, and conditionally augmenting direct response with cognitive consolidation and multi-flight external reasoning.
    \item We develop a progressive pipeline staging duplex behaviors through a three-stage Thinker curriculum by data difficulty, then Talker-only and joint SFT, plus preliminary GRPO.
\end{enumerate}

\begin{figure*}[t]
    \centering
    \includegraphics[width=\textwidth]{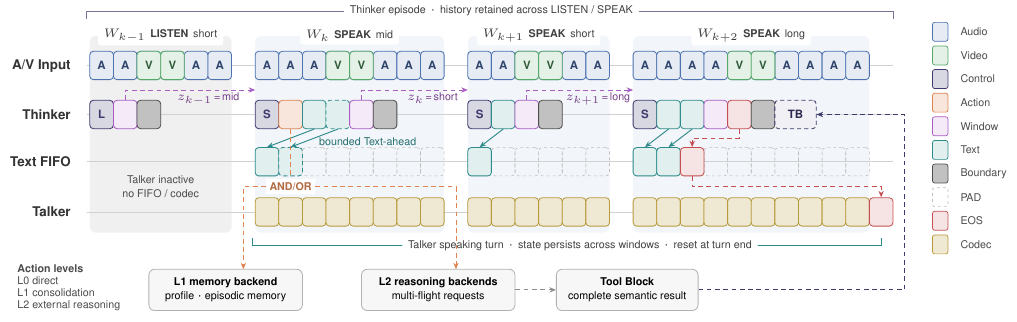}\\[1pt]
    {\scriptsize\textit{e.g., $W_{k-1}$:}\enspace\texttt{<\textbar listen\textbar>}\enspace\texttt{<\textbar window:mid\textbar>}\enspace\texttt{<\textbar chunk\_eos\textbar>};\quad\textit{$W_k$:}\enspace\texttt{<\textbar speak\textbar>}\enspace\texttt{<\textbar launch:0\textbar>}\enspace\texttt{Let me check.}\enspace\texttt{<\textbar window:short\textbar>}\enspace\texttt{<\textbar chunk\_eos\textbar>}}
    \caption{AdaptDuplex overview: one interleaved token protocol carries adaptive windows, dual-stream alignment, and tiered reasoning. Window bands, token placements, and A/V interleaving are schematic; explicit decision tokens allow inference-time control via logits bias.}
    \label{fig:overview}
\end{figure*}

\section{Proposed Method}
\label{sec:method}
AdaptDuplex inherits the Thinker--Talker architecture of Qwen3-Omni~\cite{xu2025qwen3omnitechnicalreport}, but moves the Thinker from generating a complete response per user turn to making per-window interaction decisions throughout an episode, with reply text only as an optional output.
The Talker keeps its synthesis role but changes its conditioning.

\subsection{Token-level Sequence Protocol}
\label{ssec:protocol}

Each window is organized as a \emph{canonical sequence}, enclosed by Unit delimiters.
The Thinker consumes the causally available environment and then emits, in a fixed order, a listen/speak control; a possibly empty sequence of action tokens drawn from \tok{memory} and the indexed \tok{launch:N}/\tok{cancel:N} families; optional assistant text; a next-window token; and boundary tokens.
The window that ends a speaking turn emits \tok{turn\_eos} before \tok{chunk\_eos}.
The complete grammar is given in the \firstappendixmention{app:grammar}.

The canonical sequence also drives speech generation.
When a \tok{speak} decision starts a speaking turn, the Talker prefills once from the environment history through that Unit, instead of re-collecting user-turn segments; later Units reach the Thinker but not the current Talker.
After one bootstrap, the Talker generates codec tokens autoregressively, popping one text condition per acoustic position from a turn-local FIFO that queues assistant text as the Thinker emits it, rather than consuming text as a continuous stream; \tok{tts\_pad} fills empty positions.
Queued text carries across windows but never across speaking turns.
On \tok{turn\_eos}, \tok{tts\_eos} is enqueued; after the remaining speech, the Talker predicts \tok{codec\_eos} and resets its turn-local state.

\noindent\textbf{A. Compact serialization.}
DuplexOmni~\cite{huang2026duplexomnirealtimelisteningseeing} serializes per-window metadata as JSON.
We instead encode protocol fields as compact tokens in a fixed order, reducing structural decoding overhead (\S\ref{ssec:protocol_eval}).

\noindent\textbf{B. Dual-stream alignment.}
The FIFO coordinates the Thinker and Talker despite their different token rates.
During training, bounded Text-ahead moves an order-preserving prefix of up to $d_k\le d_{\max}=2$ text tokens from each window into the preceding window of the same speaking turn.
This changes only per-window allocation: the flattened speaking-turn text sequence, control decisions, codec targets, and causal boundary for future environments remain unchanged.
With interleaved text/\tok{tts\_pad} conditioning, this yields streaming text--speech alignment with a bounded text lead without modifying the Talker architecture.

\noindent\textbf{C. Runtime controllability.}
Because behavioral decisions are explicit tokens, additive logits bias adjusts runtime behavior at inference without modifying model weights.

\subsection{Adaptive Interaction Mechanisms}
\label{ssec:mechanisms}

The canonical sequence exposes two context-dependent runtime decisions: temporal granularity and action composition.

\noindent\textbf{A. Dynamic window prediction.}
Fixed windows impose a global latency--evidence trade-off: at a turn boundary, a short window must decide on little post-boundary acoustic evidence, whereas a long window gathers more evidence but responds slowly.
At the end of window $k$, the model predicts $z_k\in\{$short, mid, long$\}$ from only causally available multimodal context, setting window $k{+}1$ to 0.48, 0.64, or 0.96~s (6, 8, or 12 codec frames at 12.5~Hz); mid is the fallback under insufficient evidence and initializes the first window.
Weak labels for $z_k$ follow priority rules over timestamps, acoustic and semantic boundaries, and overlap labels (see \appendixmention{app:window}).

\noindent\textbf{B. Conditional action composition.}
A binary choice between direct response and external reasoning overlooks an intermediate cognitive-consolidation operation.
We therefore augment direct response (L0, an empty action sequence) with two independently selectable, non-blocking action families that may co-occur within a window.
L1 emits \tok{memory} to consolidate state through a memory backend; later Units see the update without stalling the current response.
L2 manages up to four in-flight requests to a pluggable external LLM, tool, or agent via indexed \tok{launch:N}/\tok{cancel:N}, $N\in\{0,\ldots,3\}$, where $N$ names a runtime slot rather than a request category: \tok{launch:N} starts an asynchronous request in idle slot $N$, and \tok{cancel:N} cancels the request in that slot and is valid only while one exists.
Valid results and explicit backend failures re-enter the Thinker at Unit boundaries as loss-masked, semantically complete \emph{Tool Blocks}, letting the model distinguish failure from no return; cancelled, stale, or unreturned requests inject nothing, and neither a failure nor a non-return blocks the current response.

\subsection{Progressive Training Pipeline}
\label{ssec:training}

To avoid the entanglement noted in \S\ref{sec:introduction}, we structure SFT along two orthogonal axes: data difficulty for the Thinker, and trainable modules across the Thinker--Talker stack.

\noindent\textbf{A. Curriculum Thinker SFT.}
The Thinker is adapted in three stages of increasing duplex difficulty.
Stage~1 isolates the canonical protocol on turn-based conversations under a fixed middle-window schedule, without window-token supervision; Stage~2 adds basic duplex events (backchannels, short pauses, and simple interruptions) and activates dynamic window prediction; Stage~3 introduces the full duplex data with \tok{memory} and indexed \tok{launch:N}/\tok{cancel:N}, requiring it to compose interaction and external-reasoning orchestration actions.
An automated multi-agent pipeline generates training data (\appendixmention{app:training}).

\noindent\textbf{B. Talker and joint SFT.}
Talker-only SFT then freezes the Thinker and trains the Talker under FIFO text conditions with bounded Text-ahead; a final joint stage calibrates all components end to end on a smaller set of high-quality duplex conversations.

\noindent\textbf{C. Policy optimization.}
GRPO continues from the joint-SFT checkpoint as a preliminary increment, refining the interaction policy without architectural change; its reward design is in the \appendixmention{app:grpo}.

\section{Experiments}
\label{sec:experiments}
\subsection{Experimental Setup}
\label{ssec:setup}

\noindent\textbf{Model.}
We compare AdaptDuplex with the native full-duplex baselines DuplexOmni and MiniCPM-o~4.5 and with a half-duplex reference, Qwen3-Omni with VAD endpointing and barge-in cancellation. DuplexOmni shares our Qwen3-Omni base but draws on about 3.02M raw conversations~\cite{huang2026duplexomnirealtimelisteningseeing}, an estimated four times our speech hours; MiniCPM-o~4.5 builds on the dense Qwen3-8B, with fewer total but more active parameters than our 30B-A3B MoE (3B active). All three duplex systems stream on a single H100 GPU (baselines with official code and parameters), so latencies include computation; since Thinker, Talker, and Code2Wav run in parallel, speech starts once the first text token is decoded, without waiting for the rest of the window's output. AdaptDuplex is initialized from Qwen3-Omni-30B-A3B-Instruct and trained on 8 H100 GPUs. Bounded Text-ahead uses three lead schedules: Aligned ($d_k{=}0$, forced-alignment windows without look-ahead), Fixed Ahead (all internal leads at $d_{\max}$), and Random Ahead (a bounded random walk over $d_k$). Unless stated otherwise, all variants before joint SFT share one Talker from Talker-only SFT with the Random Ahead schedule; Table~\ref{tab:protocol} adapts one Talker per schedule. Decoding follows the official Qwen3-Omni policy (greedy Thinker, sampled Talker).

\noindent\textbf{Data.}
Curriculum SFT and its single-stage ablation, Mixed SFT, use an internally constructed synthetic full-duplex corpus of 765,424 dialogues (17,259 h, of which 10,857 h are assistant speech); Mixed SFT matches Curriculum in instance count and per-instance exposure on this corpus. Joint SFT continues from the Curriculum checkpoint on a smaller high-quality subset of DuplexDrama~\cite{guo2026duplexdramasynthesizeddialoguedataset}.

\noindent\textbf{Evaluation.}
We use Full-Duplex-Bench (FDB)~\cite{lin2025fullduplexbenchbenchmarkevaluatefullduplex}, its v1.5 extension (FDB-v1.5)~\cite{lin2026fullduplexbenchv15evaluatingoverlap} with genuine interruptions and other overlap scenarios, and the human-recorded, scripted Chinese and English HumDial-FDBench~\cite{wang2026fullduplexinteractionspokendialogue}. Both FDB evaluations follow their official scoring and transcription protocols. Serialization and replay share 24 streaming records; replay re-segments them under each fixed duration and the dynamic policy. Text-ahead conditions share utterances, Thinker text, and window schedules; naturalness uses the official UTMOSv2~\cite{baba2024t05voicemoschallenge2024}. L1/L2 action prediction is evaluated from gold prefixes, with positive-class F1 for the L1 memory action and action EM, token-level launch/cancel F1, and invalid-action rate for L2; memory instances and L2 scenarios from the same episodes use gold context. No FDB or HumDial scripts, audio, or speakers enter data generation, no setting is tuned on test sets, and benchmarking attaches no L1/L2 backend or DuplexOmni thinking layer.

\subsection{Overall Full-Duplex Performance}
\label{ssec:main_results}

\begin{table}[!t]
\caption{Full-duplex interaction performance on (a) FDB-v1 turn-taking and (b) FDB-v1.5 overlap handling.}
\label{tab:fdb}
\begingroup
\ifdefined\tablepreviewvalues
\newcommand{\tablevalue}[2]{#1}
\else
\newcommand{\tablevalue}[2]{#2}
\fi
\newcommand{\modelhead}[1]{\setbox0=\hbox{00.000}\setbox1=\hbox{\begin{tabular}[c]{@{}c@{}}#1\end{tabular}}\ifdim\wd1>\wd0\box1\else\makebox[\wd0][c]{\box1}\fi}
\centering
\footnotesize
\setlength{\tabcolsep}{1.5pt}
\setlength{\aboverulesep}{0.25ex}
\setlength{\belowrulesep}{0.25ex}
\makebox[\linewidth][l]{\hspace{0.696pt}\begin{tabular}{@{}>{\raggedright\arraybackslash}m{29.45698pt}>{\raggedright\arraybackslash}m{37.59372pt}>{\centering\arraybackslash}m{24.39218pt}>{\centering\arraybackslash}m{23.98372pt}>{\centering\arraybackslash}m{25.76787pt}>{\centering\arraybackslash}m{32.45558pt}>{\centering\arraybackslash}m{23.43193pt}>{\centering\arraybackslash}m{23.43193pt}@{}}
\toprule
\multicolumn{1}{@{}l}{Task} & \multicolumn{1}{l}{Metric} & \multicolumn{1}{c}{\modelhead{Moshi$^\dagger$}} & \multicolumn{1}{c}{\modelhead{Freeze-\\Omni$^\dagger$}} & \multicolumn{1}{c}{\modelhead{Qwen3-\\Omni\\+VAD}} & \multicolumn{1}{c}{\modelhead{MiniCPM\\-o 4.5}} & \multicolumn{1}{c}{\modelhead{Duplex\\Omni}} & \multicolumn{1}{c@{}}{\modelhead{Adapt\\Duplex}} \\
\midrule
\multicolumn{8}{@{}l}{\textbf{(a) FDB-v1: turn-taking}} \\
\midrule
\multirow{2}{*}{Pause} & \mbox{Syn. TOR$\downarrow$} & 0.985 & 0.642 & 0.993 & 0.270 & \textbf{0.116} & \tablevalue{0.148}{0.128} \\
& \mbox{Can. TOR$\downarrow$} & 0.980 & 0.481 & 0.949 & 0.463 & \textbf{0.097} & \tablevalue{0.412}{0.299} \\
\midrule
\multirow{3}{*}{\shortstack[l]{Back-\\channel}} & TOR$\downarrow$ & 1.000 & 0.636 & 0.344 & \textbf{0.273} & 0.491 & \tablevalue{0.327}{0.334} \\
& Freq.$\uparrow$ & 0.001 & 0.001 & 0.018 & 0.020 & 0.034 & \tablevalue{\textbf{0.035}}{\textbf{0.040}} \\
& JSD$\downarrow$ & 0.957 & 0.997 & 0.879 & 0.902 & 0.847 & \tablevalue{0.863}{\textbf{0.821}} \\
\midrule
\multirow{2}{*}{\shortstack[l]{Smooth\\turn}} & TOR$\uparrow$ & 0.941 & 0.336 & \textbf{0.992} & 0.882 & 0.193 & \tablevalue{0.723}{0.982} \\
& Resp. lat.$\downarrow$ & 0.265 & 0.953 & 3.358 & 1.222 & 0.000$^\ddagger$ & \tablevalue{\textbf{0.147}}{\textbf{0.085}} \\
\midrule
\multirow{3}{*}{\shortstack[l]{Inter-\\ruption}} & TOR$\uparrow$ & \textbf{1.000} & 0.867 & 0.995 & 0.985 & 0.746 & \tablevalue{0.940}{0.985} \\
& GPT-4o$\uparrow$ & 0.765 & 3.615 & \textbf{4.824} & 4.497 & 4.222 & \tablevalue{3.239}{4.512} \\
& Resp. lat.$\downarrow$ & 0.257 & 1.409 & 3.380 & 1.408 & 0.158 & \tablevalue{0.334}{\textbf{0.131}} \\
\midrule
\end{tabular}}\par\nointerlineskip
\makebox[\linewidth][l]{\hspace{0.696pt}\begin{tabular}{@{}>{\raggedright\arraybackslash}m{29.45698pt}>{\raggedright\arraybackslash}m{37.59372pt}>{\centering\arraybackslash}m{24.39218pt}>{\centering\arraybackslash}m{23.98372pt}>{\centering\arraybackslash}m{25.76787pt}>{\centering\arraybackslash}m{32.45558pt}>{\centering\arraybackslash}m{23.43193pt}>{\centering\arraybackslash}m{23.43193pt}@{}}
\multicolumn{8}{@{}l}{\textbf{(b) FDB-v1.5: overlap handling}} \\
\midrule
\multirow{3}{*}{\shortstack[l]{Inter-\\ruption}} & Behavior$\uparrow$ & 0.50 & 0.72 & \textbf{0.97} & 0.82 & 0.67 & \tablevalue{0.62}{0.79} \\
& Stop lat.$\downarrow$ & 1.16 & 1.42 & \textbf{0.09} & 2.49 & 2.01 & \tablevalue{2.86}{1.32} \\
& Resp. lat.$\downarrow$ & 1.47 & 1.35 & 3.70 & 1.86 & \tablevalue{\textbf{0.36}}{0.36} & \tablevalue{1.25}{\textbf{0.34}} \\
\midrule
\multirow{3}{*}{\shortstack[l]{Back-\\channel}} & Behavior$\uparrow$ & 0.06 & \textbf{0.80} & 0.00 & 0.46 & 0.28 & \tablevalue{0.41}{0.55} \\
& Stop lat.$\uparrow$ & 0.42 & 0.66 & -- & 0.76 & 0.80 & \tablevalue{0.76}{\textbf{0.84}} \\
& Resp. lat.$\downarrow$ & 3.00 & 2.16 & 2.73 & 2.34 & \tablevalue{\textbf{0.82}}{0.82} & \tablevalue{1.77}{\textbf{0.43}} \\
\midrule
\multirow{3}{*}{\shortstack[l]{Talking\\to others}} & Behavior$\uparrow$ & 0.20 & 0.25 & 0.01 & 0.21 & \textbf{0.37} & \tablevalue{0.28}{0.35} \\
& Stop lat.$\uparrow$ & 0.87 & 1.39 & 0.10 & 1.71 & 1.87 & \tablevalue{\textbf{1.89}}{\textbf{1.89}} \\
& Resp. lat.$\downarrow$ & 2.38 & \tablevalue{\textbf{1.32}}{1.32} & 3.16 & 1.85 & 1.95 & \tablevalue{1.53}{\textbf{0.33}} \\
\midrule
\multirow{3}{*}{\shortstack[l]{Back-\\ground\\speech}} & Behavior$\uparrow$ & 0.08 & 0.26 & 0.00 & 0.37 & 0.07 & \tablevalue{0.24}{\textbf{0.43}} \\
& Stop lat.$\uparrow$ & 0.54 & 0.98 & 0.09 & 1.46 & 1.50 & \tablevalue{1.46}{\textbf{1.51}} \\
& Resp. lat.$\downarrow$ & 1.62 & 1.60 & 3.19 & 1.61 & \tablevalue{\textbf{0.74}}{0.74} & \tablevalue{2.07}{\textbf{0.36}} \\
\bottomrule
\end{tabular}}

\par\vspace{3pt}
\raggedright\footnotesize
(a): Syn./Can.: synthetic/Candor pauses; Freq. in events/s; GPT-4o scores (0--5) and latencies (s) conditional on takeover (TO=1); negative per-sample latencies (annotated turn ends lag true ones) clipped to 0. $^\ddagger$All samples negative; excluded from highlighting. (b): Behavior is 1 for Respond (interruption) or Resume (other scenarios), 0.5 for Uncertain (talking-to-others, background-speech), 0 otherwise; --: no qualifying event. $^\dagger$Published results (FDB~\cite{lin2025fullduplexbenchbenchmarkevaluatefullduplex} Tab.~III; FDB-v1.5~\cite{lin2026fullduplexbenchv15evaluatingoverlap} Tab.~2); latencies not hardware-matched. Abstract win counts exclude $^\ddagger$ and treat ties as non-wins.
\par
\endgroup

\end{table}

Table~\ref{tab:fdb} isolates scene judgment on the turn-taking and overlap cases: versus DuplexOmni, AdaptDuplex more often withholds a full turn when only a backchannel is due (FDB-v1 takeover rate, TOR, 0.334 vs.~0.491; MiniCPM-o~4.5 is lowest at 0.273) and resumes more reliably under backchannel and background-speech overlap (FDB-v1.5 behavior 0.55 vs.~0.28; 0.43 vs.~0.07). Its higher Candor pause TOR (0.299 vs.~0.097) is the flip side of taking the floor when due (smooth-turn TOR 0.982 vs.~0.193). These gains do not cost speed: response latency is sub-second in all reported FDB scenarios and the lowest among native duplex systems in every FDB-v1.5 scenario, and on FDB-v1 interruptions AdaptDuplex is both more accurate (TOR 0.985 vs.~0.746) and faster (0.131 vs.~0.158~s); MiniCPM-o~4.5 matches that TOR at 1.408~s.

\begin{table}[!t]
\caption{HumDial-FDBench evaluation on Chinese (ZH) and English (EN) human-recorded interactions.}
\label{tab:humdial}
\begingroup
\ifdefined\tablepreviewvalues
\newcommand{\tablevalue}[2]{#1}
\else
\newcommand{\tablevalue}[2]{#2}
\fi
\newcommand{\modelhead}[1]{\setbox0=\hbox{00.000}\setbox1=\hbox{\begin{tabular}[c]{@{}c@{}}#1\end{tabular}}\ifdim\wd1>\wd0\box1\else\makebox[\wd0][c]{\box1}\fi}
\centering
\footnotesize
\setlength{\tabcolsep}{1.5pt}
\setlength{\aboverulesep}{0.25ex}
\setlength{\belowrulesep}{0.25ex}
\begin{tabular}{@{}>{\raggedright\arraybackslash}p{42pt}lcccccc@{}}
\toprule
Metric & Split & \modelhead{Moshi$^\dagger$} & \modelhead{Freeze-\\Omni$^\dagger$} & \modelhead{Qwen3-\\Omni\\+VAD} & \modelhead{MiniCPM\\-o 4.5} & \modelhead{Duplex\\Omni} & \modelhead{Adapt\\Duplex} \\
\midrule
\multirow{3}{*}{\mbox{Int.$\uparrow$ (\%)}} & ZH & NR & NR & \textbf{91.3} & 81.8 & 56.4 & \tablevalue{89.6}{90.4} \\
& EN & NR & NR & \textbf{80.9} & 63.0 & 55.2 & \tablevalue{73.8}{74.4} \\
& ALL & 35.4 & 29.6 & \textbf{86.1} & 72.4 & 55.8 & \tablevalue{81.7}{82.2} \\
\midrule
\multirow{3}{*}{\mbox{Rej.$\uparrow$ (\%)}} & ZH & NR & NR & 36.8 & 44.4 & 40.0 & \tablevalue{\textbf{46.9}}{\textbf{51.8}} \\
& EN & NR & NR & 34.2 & \textbf{49.6} & 47.8 & \tablevalue{43.5}{46.1} \\
& ALL & 22.8 & \textbf{50.2} & 35.6 & 47.2 & 44.1 & \tablevalue{45.0}{49.1} \\
\midrule
\multirow{3}{*}{\mbox{Delay$\downarrow$ (s)}} & ZH & NR & NR & 4.670 & 2.074 & 4.266 & \tablevalue{\textbf{0.912}}{\textbf{0.835}} \\
& EN & NR & NR & 4.924 & 2.315 & 4.590 & \tablevalue{\textbf{1.287}}{\textbf{1.345}} \\
& ALL & 2.876 & 2.578 & 4.771 & 2.203 & 4.459 & \tablevalue{\textbf{1.064}}{\textbf{1.042}} \\
\midrule
D-Sco.$\uparrow$ & ALL & 56.3 & 59.5 & 41.9 & 64.0 & 43.8 & \tablevalue{\textbf{99.6}}{\textbf{85.4}} \\
Final$\uparrow$ & ALL & 34.5 & 43.8 & 57.1 & 60.6 & 48.7 & \tablevalue{\textbf{70.6}}{\textbf{69.6}} \\
\bottomrule
\end{tabular}
\par\vspace{3pt}
\raggedright\footnotesize
NR: not reported. ALL pools ZH and EN under the official protocol. Int./Rej.: averaged subscenario scores; Delay: combined applicable latency types. D-Sco.: official logarithmic normalization of ALL Delay; Final: 0.4 Int. + 0.4 Rej. + 0.2 D-Sco. $^\dagger$From HumDial-FDBench~\cite{wang2026fullduplexinteractionspokendialogue} Tab.~2.
\par
\endgroup

\end{table}

On human-recorded interactions with the same operating point (Table~\ref{tab:humdial}), AdaptDuplex attains the highest Final score among the compared duplex dialogue models (69.6), with delays of 0.835~s (ZH) and 1.345~s (EN), faster than MiniCPM-o~4.5 and DuplexOmni. The half-duplex Qwen3-Omni+VAD remains more accurate at interruption (86.1 vs.~82.2 ALL), whereas AdaptDuplex is more accurate at rejection (49.1 vs.~35.6 ALL).

\subsection{Protocol Efficiency, Alignment, and Control}
\label{ssec:protocol_eval}

\begin{table}[t]
\caption{Talker text--speech alignment under the three Text-ahead lead schedules (1,516 ZH + 1,514 EN paired utterances).}
\label{tab:protocol}
\begingroup
\ifdefined\tablepreviewvalues
\newcommand{\tablevalue}[2]{#1}
\else
\newcommand{\tablevalue}[2]{#2}
\fi
\centering
\footnotesize
\setlength{\tabcolsep}{0pt}
\setlength{\aboverulesep}{0.25ex}
\setlength{\belowrulesep}{0.25ex}
\begin{tabular}{@{}m{0.205\linewidth}>{\centering\arraybackslash}m{0.125\linewidth}>{\centering\arraybackslash}m{0.155\linewidth}>{\centering\arraybackslash}m{0.125\linewidth}>{\centering\arraybackslash}m{0.13\linewidth}>{\centering\arraybackslash}m{0.13\linewidth}>{\centering\arraybackslash}m{0.13\linewidth}@{}}
\toprule
Training & \shortstack{UTMOS\\v2$\uparrow$} & \shortstack{CER/WER\\$\downarrow$(\%)} & \shortstack{ValidEnd\\$\uparrow$(\%)} & \shortstack{EarlyCut\\$\downarrow$(\%)} & \shortstack{Late5+\\$\downarrow$(\%)} & \shortstack{NoEOS\\$\downarrow$(\%)}\\
\midrule
Aligned & \tablevalue{3.68}{2.90} & \tablevalue{8.42/6.85}{10.74/19.10} & \tablevalue{72.50}{50.00} & \tablevalue{4.50}{0.00} & \tablevalue{14.00}{8.02} & \tablevalue{4.00}{0.36}\\
\mbox{Fixed Ahead} & \tablevalue{3.94}{3.55} & \tablevalue{5.76/4.32}{5.38/10.03} & \tablevalue{82.00}{84.75} & \tablevalue{3.00}{0.00} & \tablevalue{8.00}{1.85} & \tablevalue{2.00}{0.07}\\
\mbox{Random Ahead} & \tablevalue{3.96}{3.65} & \tablevalue{5.21/3.98}{4.83/8.93} & \tablevalue{91.50}{86.37} & \tablevalue{2.00}{0.00} & \tablevalue{2.50}{1.49} & \tablevalue{0.50}{0.03}\\
\bottomrule
\end{tabular}
\par\vspace{3pt}
\raggedright\footnotesize
Entries: ZH CER / EN WER (\%, Qwen3-ASR-1.7B). EOS offsets in codec positions. ValidEnd: complete by final-window boundary; EarlyCut: truncated; Late5+: $[5,16)$; NoEOS: $[16,+\infty)$; rest: $[1,5)$.
\par
\endgroup

\end{table}

To test serialization, we hold the semantic records, the fixed 0.48-s window, and the adapted Talker constant and replace only the canonical sequence with a matched DuplexOmni-style JSON encoding of the same fields on 3,724 paired windows. Canonical serialization cuts tokens per window from 26.440 to 5.867 ($-78\%$) and the protocol's share of output tokens from 93.88\% to 68.70\%, removes structurally invalid windows (0.48\% to 0.00\%), and lowers the per-window GPU busy time (union of Thinker, Talker, and Code2Wav) at P50/P90/P99 from 789.4/957.6/1,164.2~ms to 418.5/491.2/672.2~ms, i.e., by 47\%/49\%/42\% (percentiles conditional on valid completion).

Table~\ref{tab:protocol} tests dual-stream alignment, varying only the Text-ahead lead schedule. Any look-ahead helps sharply over Aligned: Fixed Ahead and Random Ahead raise pooled UTMOSv2 from 2.90 to 3.55 and 3.65, roughly halve bilingual CER/WER, and lift ValidEnd from 50.00\% to 84.75\% and 86.37\%. Random Ahead, the default Talker schedule, is best on naturalness, error, and ValidEnd.

Table~\ref{tab:ablation}(b) tests runtime control by adding $\ln\alpha$ to the speak logit ($\alpha\in\{0.5,1,2\}$; $\times$1 is the untuned policy) at the Curriculum checkpoint, with the Thinker, Talker, window policy, and test inputs fixed. Raising $\alpha$ from 0.5 to 2 moves the operating point monotonically: interruption behavior rises (0.76$\to$0.81) and response latency shortens (0.42$\to$0.33~s), while backchannel, talking-to-others, and background-speech scores fall and yield slows slightly (1.28$\to$1.36~s)---an inference-time listen/speak trade-off.

\begin{table}[t]
\caption{Controlled variations of the default system (Curriculum SFT checkpoint, dynamic window, no logit bias) on FDB-v1.5. (a) and (b) change one inference-time setting of the same checkpoint; (c) changes the training recipe.}
\label{tab:ablation}
\begingroup
\ifdefined\tablepreviewvalues
\newcommand{\tablevalue}[2]{#1}
\else
\newcommand{\tablevalue}[2]{#2}
\fi
\centering
\footnotesize
\setlength{\tabcolsep}{0pt}
\setlength{\aboverulesep}{0.25ex}
\setlength{\belowrulesep}{0.25ex}
\begin{tabular}{@{}>{\raggedright\arraybackslash}m{0.27\linewidth}*{7}{>{\centering\arraybackslash}m{0.1042857\linewidth}}@{}}
\toprule
Variant & \multicolumn{7}{@{}c@{}}{%
\begin{tabular}[c]{@{}*{7}{>{\centering\arraybackslash}m{0.1042857\linewidth}}@{}}
\multicolumn{4}{c}{Behavior} & \multicolumn{3}{c}{Timing (s)} \\
\cmidrule(lr){1-4}\cmidrule(l){5-7}
Int.$\uparrow$ & Back.$\uparrow$ & Talk.$\uparrow$ & Bg.$\uparrow$ & Yield$\downarrow$ & Hold$\uparrow$ & Resp.$\downarrow$ \\
\end{tabular}}\\
\midrule
Default & \tablevalue{0.85}{0.79} & \tablevalue{0.92}{0.55} & \tablevalue{0.86}{0.35} & \tablevalue{0.88}{0.43} & \tablevalue{0.60}{1.32} & \tablevalue{0.79}{1.41} & \tablevalue{1.50}{0.37} \\
\midrule
\multicolumn{8}{@{}l}{\emph{(a) Window policy}}\\
\quad Fixed 0.48\,s & \tablevalue{0.84}{0.82} & \tablevalue{0.87}{0.52} & \tablevalue{0.80}{0.31} & \tablevalue{0.82}{0.34} & \tablevalue{0.58}{1.19} & \tablevalue{0.62}{1.38} & \tablevalue{1.65}{0.32} \\
\quad Fixed 0.64\,s & \tablevalue{0.83}{0.78} & \tablevalue{0.90}{0.54} & \tablevalue{0.82}{0.33} & \tablevalue{0.84}{0.43} & \tablevalue{0.67}{1.81} & \tablevalue{0.74}{1.40} & \tablevalue{1.58}{0.45} \\
\quad Fixed 0.96\,s & \tablevalue{0.79}{0.76} & \tablevalue{0.91}{0.56} & \tablevalue{0.86}{0.34} & \tablevalue{0.87}{0.46} & \tablevalue{0.82}{2.61} & \tablevalue{0.88}{1.46} & \tablevalue{1.54}{0.66} \\
\midrule
\multicolumn{8}{@{}l}{\emph{(b) Speak-logit scale $\alpha$}}\\
\quad $\times$0.5 & \tablevalue{0.76}{0.76} & \tablevalue{0.58}{0.58} & \tablevalue{0.26}{0.38} & \tablevalue{0.45}{0.46} & \tablevalue{1.28}{1.28} & \tablevalue{1.26}{1.34} & \tablevalue{0.42}{0.42} \\
\quad $\times$2.0 & \tablevalue{0.81}{0.81} & \tablevalue{0.49}{0.49} & \tablevalue{0.20}{0.32} & \tablevalue{0.36}{0.37} & \tablevalue{1.36}{1.36} & \tablevalue{1.40}{1.48} & \tablevalue{0.33}{0.33} \\
\midrule
\multicolumn{8}{@{}l}{\emph{(c) Training recipe}}\\
\quad Mixed SFT & \tablevalue{0.54}{0.62} & \tablevalue{0.35}{0.41} & \tablevalue{0.22}{0.28} & \tablevalue{0.19}{0.24} & \tablevalue{3.02}{2.86} & \tablevalue{1.26}{1.40} & \tablevalue{1.81}{1.65} \\
\quad Default\,+\,Joint & \tablevalue{0.62}{0.83} & \tablevalue{0.41}{0.55} & \tablevalue{0.28}{0.37} & \tablevalue{0.24}{0.46} & \tablevalue{2.86}{1.20} & \tablevalue{1.37}{1.44} & \tablevalue{1.66}{0.32} \\
\quad \quad $+$\,GRPO & \tablevalue{0.65}{0.85} & \tablevalue{0.44}{0.57} & \tablevalue{0.31}{0.44} & \tablevalue{0.27}{0.45} & \tablevalue{2.71}{1.08} & \tablevalue{1.40}{1.45} & \tablevalue{1.58}{0.30} \\
\bottomrule
\end{tabular}
\par\vspace{3pt}
\raggedright\footnotesize
Behavior uses the scenario mapping of Table~\ref{tab:fdb}(b). Yield = interruption stop latency; Hold = macro-averaged stop latency over the three non-interruption scenarios; Resp. = macro-averaged response latency over all four scenarios; both macro-averages use equal scenario weights. All rows: free-running end-to-end evaluation. (b): additive $\ln\alpha$ bias on the speak logit ($\times$1 $=$ Default, no bias). (c): Mixed SFT is the matched-exposure single-stage control; $+$\,GRPO continues from Default\,+\,Joint.
\par
\endgroup

\end{table}

\subsection{Adaptive Interaction}
\label{ssec:adaptive_eval}

\noindent\textbf{Dynamic window.}
Table~\ref{tab:ablation}(a) runs the Default checkpoint with each window duration fixed at inference via window-token bias and makes the fixed-window dilemma explicit: the short window is fastest but lowest on backchannel, whereas the long window is best on backchannel and background speech but slowest, and its longest hold comes with the slowest yield. The dynamic policy stays within reach of each extreme at once---near-short latency (0.37~s, 1.32~s) with near-long backchannel (0.55)---and is the single best setting on talking-to-others (0.35). Controlled autoregressive replay shows the policy is not degenerate: at 43.9~GPU-s/min (busy time of one H100 per input minute, i.e., $0.73\times$ real time) it sits between always-short (48.3) and always-long (33.5), so it is not defaulting to the longest, cheapest window. The SPEAK/LISTEN breakdown, per-stage results, and window-policy matrices are in the \appendixmention{app:experiments}.

\noindent\textbf{Tiered actions.}
From gold prefixes, the model predicts the sparse L1 \tok{memory} action at F1 80.66\% and L2 orchestration at action EM 76.84\% (exact match of the executed action set) and launch F1 68.63\%, with cancellation the weakest sub-decision (F1 54.06\%); concurrency is the main stressor, with EM falling to 65.17\% and the invalid-action rate nearly doubling (3.80\%$\to$7.10\%) when at least two requests are in flight. With gold PROFILE and MEMORY context, L1 fact accuracy rises from 10.25\% to 62.75\%, so the Thinker uses injected consolidated state. For L2, canary accuracy with gold results (all target and no forbidden canaries in the Thinker text) is 86.50/73.50/68.50\% for single, distracted, and composed queries, and 81.50/80.50\% under injected failure and silent non-return.

\subsection{Training Strategy}
\label{ssec:training_eval}

Table~\ref{tab:ablation}(c) compares training recipes, all with the dynamic window policy; Mixed SFT is the matched-exposure single-stage control. Curriculum accounts for most of the gain over Mixed SFT, most visibly on latency (response 1.65$\to$0.37~s), and improves behavior in all four scenarios. Joint SFT (from Curriculum) and GRPO (from Joint) add mostly monotonic changes, clearest for talking-to-others (0.35$\to$0.44), yield, and response; backchannel is unchanged after Joint SFT, and background-speech dips slightly after GRPO. Other results, including Tables~\ref{tab:fdb}--\ref{tab:humdial}, use the Curriculum checkpoint, which Joint SFT matches or improves on every Table~\ref{tab:ablation} metric; further protocol-cost and L1/L2 analyses are in the \appendixmention{app:experiments}.

\section{Conclusion}
\label{sec:conclusion}
We presented AdaptDuplex, which extends Qwen3-Omni by co-designing a compact token-level protocol, adaptive window and action mechanisms, and a progressive training pipeline.

Several limitations remain: evaluation currently covers Chinese and English only, the window controller selects among three discrete durations rather than a continuous horizon, and external reasoning is bounded to four concurrent requests. General-ability retention after duplex adaptation and end-to-end L1/L2 evaluation with real backends remain to be measured. Future work will study continuous window prediction, more languages, adaptive concurrency limits, training and evaluation with visual inputs, and multi-party conversation.

\section{Acknowledgment}
\label{sec:acknowledgment}
The authors used Claude (Anthropic, Claude Sonnet 5) to improve the language and
grammar of the manuscript and to obtain suggestions on the organization of
Sections 1 and 2. All technical content, experiments, and conclusions were produced
by the authors, who reviewed and take full responsibility for the final text.

\bibliographystyle{IEEEbib}
\bibliography{refs}

\ifdefined\extendedversion
  \clearpage
  \appendix
  
\fi
\fi

\end{document}